%% file: main.tex
\documentclass[review]{elsarticle}
\usepackage[margin=1in]{geometry}
\usepackage{lineno,hyperref}
\modulolinenumbers[5]
\usepackage{float}
\usepackage{amssymb}
\usepackage{graphicx}
\usepackage{multicol}
\usepackage{multirow}
\usepackage{nomencl}
\usepackage{amsmath}
\usepackage{todonotes}
\usepackage{comment}
\usepackage{subcaption,booktabs}
\usepackage{atbegshi}
\AtBeginDocument{\AtBeginShipoutNext{\AtBeginShipoutDiscard}}

\biboptions{authoryear}

\begin{document}

\begin{frontmatter}

\title{A Decision Support Framework for Optimal Vaccine Distribution Across a Multi-tier Cold Chain Network}




\author[1]{Shanmukhi Sripada}
\author[1]{Ayush Jain}
\author[2]{Prasanna Ramamoorthy}
\author[1]{Varun Ramamohan\corref{cor}}
\cormark[cor]
\ead{varunr@mech.iitd.ac.in}
\cortext[cor]{Corresponding author}

\address[1]{Department of Mechanical Engineering, Indian Institute of Technology Delhi, Hauz Khas, New Delhi 110016, India}
\address[2]{Department of Management Studies, Indian Institute of Technology Delhi, Hauz Khas, New Delhi 110016, India}

\begin{abstract}
The importance of vaccination and the logistics involved in the procurement, storage and distribution of vaccines across their cold chain has come to the forefront during the COVID-19 pandemic. In this paper, we present a decision support framework for optimizing multiple aspects of vaccine distribution across a multitier cold chain network. We propose two multi-period optimization formulations within this framework: first to minimize inventory, ordering, transportation, personnel and shortage costs associated with a single vaccine; the second being an extension of the first for the case when multiple vaccines with differing efficacies and costs are available for the same disease. Vaccine transportation and administration lead times are also incorporated within the models. We also develop robust optimization versions of the single vaccine model to account for the impact of uncertainty in model parameters on the optimal vaccine distribution solution. We use the case of the Indian state of Bihar and COVID-19 vaccines to illustrate the implementation of the framework. We present computational experiments to demonstrate: (a) the organization of the model outputs; (b) how the models can be used to assess the impact of cold chain point storage capacities, transportation vehicle capacities, and manufacturer capacities on the optimal vaccine distribution pattern; and (c) the impact of vaccine efficacies and associated costs such as ordering and transportation costs on the vaccine selection decision informed by the model. We then consider the computational expense of the framework for realistic problem instances, and suggest multiple preprocessing techniques to reduce their computational burden. Finally, we also demonstrate how the robust versions of the single vaccine model outperform the deterministic version under multiple levels of uncertainty in key model parameters. Our study presents public health authorities and other stakeholders with a vaccine distribution and capacity planning tool for multi-tier cold chain networks.
\end{abstract}


\end{frontmatter}
\newpage

\section{Introduction}
\label{section:introduction}
\input{Sections/Introduction}
\section{Literature Review}
\label{section:literature}
\input{Sections/Literature_Review}

\section{Development of the Decision Support Framework}
\label{section:methods}
In this section, we first describe the decision problem that our framework addresses, and then describe the mathematical models within our framework. We then outline the model parameter estimation process to conclude this section.
\subsection{Decision Problem Description}
\label{subsection:problem_description}
\input{Sections/Problem_Description}
\subsection{Single Vaccine Model}
\label{subsection:single_vaccine_formulation}
\input{Sections/Mathematical_Formulation_Single_Commodity}
The extension of the single vaccine formulation to consider multiple vaccines for the same disease is provided in \ref{subsection:multiple_vaccine_formulation}.
\subsection{Robust Formulation: Single Vaccine Model}
\label{subsection:robust_vaccine_formulation}
\input{Sections/robust_methodology}

\subsection{Model Parameter Estimation}
\label{subsection:model_parametrization}
\input{Sections/Estimation_of_parameters}
\section{Computational Implementation of the Decision Support Framework}
\label{section:numerical_results}
We organize our presentation of the numerical experiments that we perform with the single vaccine, multiple vaccine and robust formulations as follows. First, we demonstrate how solving the single and multiple vaccine formulations yield results that can inform decisions associated with the distribution of vaccines across the CCP that we consider. As part of this, we consider the impact of certain parameters on the optimal ordering and inventory patterns generated by these formulations. Next, we consider the computational cost of the single vaccine formulation, and discuss preprocessing techniques that can be used to speed up solution generation as well as improve the quality of the solutions. Finally, we discuss how the robust formulation can be used and its performance with respect to the standard single vaccine formulation.

\subsection{Single Vaccine Model}
\label{subsection:single_vaccine_model}
\input{Sections/Single_commodity_results}

\subsection{Single Vaccine Model: Runtime Analysis and Preprocessing}
\label{subsection:pre_processing}
\input{Sections/Pre_processing_results}






\bibliography{mybibfile}

\newpage
\appendix
\section{Literature Review: Vaccine Composition \& Allocation}
\label{aplitreview}
We discuss the literature associated with vaccine composition and allocation below. The relevant vaccine production and distribution literature is provided in Section~\ref{section:literature}.

\subsection{Composition of Vaccines}
In the context of composition of vaccines, policymakers are interested in determining the vaccines to be used for a particular disease, since a disease might have multiple vaccines. Since this decision has to be taken before the onset of the disease due to long production times of vaccines, there is substantial uncertainty regarding demand for the vaccine and the schedule of vaccination. Researchers have studied multiple strategies to take better decisions to aid decision makers in this scenario. \cite{wu2007} suggests that including only the predicted strain of influenza for vaccine development is only slightly suboptimal. \cite{kornish2008} study the ``commit or defer" policy, in which the vaccine manufacturer at any time period commits to produce or defers the production decision until more information about the incoming demand is available. \cite{cho2010} extended the work of \cite{kornish2008} by incorporating production yield of vaccines and efficacy of vaccines in their study. \cite{ozaltin2011} studied the possibility of choosing multiple strains for inclusion in the vaccine. The authors present a multistage stochastic mixed integer program to address the problem. 

A closely related problem to the composition of vaccines is that of vial size determination. Vaccine vial sizes plays a major role in avoiding vaccine wastage, because a vial has multiple doses and once opened the vial has to be used within a specific time period. \cite{azadi2020optimization} present a case study from Niger and Nigeria, in which they study the effect of  parameters such as the number of supply chain echelons, vaccine vial size and vaccine technologies on coverage rates and vaccine availability in clinics. One limitation of the study is that operational costs have not been included. \cite{azadi2019developing} present a case study from Bangladesh in which the authors are interested in minimization of inventory replenishment and open vial wastage costs. The authors developed a two-stage stochastic programming model to address the problem. The authors show that a combination of vial sizes is more effective in reducing open vial wastage than the prevalent practice of using single-sized multi-dose vials. \cite{dhamodharan2012determining} considered both vaccine vial size determination and inventory management in which they studied optimal vaccine vial size determination and inventory re-ordering points to devise a policy that provides best trade-off between procurement costs and coverage levels.  

\subsection{Allocation of Vaccines}
Allocation of vaccines across the target population is also a problem of significant complexity. Here, the decision maker is interested in deciding which population subgroup should be vaccinated and which group should not be. Another variant of the problem also deals with allocating resources to different regions or different stakeholders to mitigate the spread of the disease. \cite{sun2009} studied a multi-player version of the problem in which different countries have their own vaccine stockpiles and at the onset of the pandemic, they have to decide how best to allocate the stock within their population and with other countries. The authors identify the conditions under which the optimal solution of a central planner (WHO) is optimal to all countries. Such a solution involves agreeing to an allocation scheme that will benefit every stakeholder. \cite{mamani2013} studied the problem of vaccine allocation between countries in which there are both international as well as intranational transmissions of the disease. The authors propose a contract which reduces both the financial burden and also the number of infected people due to the influenza pandemic. 

Under a single decision maker setting, \cite{uribe2011} studied a simulation optimization model, parameterized using historical data from a pandemic outbreak, to devise dynamic mitigation strategies to contain an influenza pandemic. \cite{yarmand2014} study a vaccine allocation model in which vaccine allocation is done in two phases. In the first phase, a limited number of doses are allocated to each region, while in the second phase additional doses are allocated so as to contain the pandemic efficiently. The authors present a two phase stochastic linear program to solve the problem.

\section{Multiple Vaccine Model}
\label{subsection:multiple_vaccine_formulation}
\input{Sections/Mathematical_Formulation_Multi_Commodity}

\section{Robust Single Vaccine Formulation with Budgeted Uncertainty Sets}
\label{aprobust}

We present the robust formulation (from Section~\ref{subsection:robust_vaccine_formulation}) of the single vaccine formulation with budgeted uncertainty sets here.

\textbf{Budgeted Uncertainty Set Formulation.}
    Budgeted uncertainty set is a less conservative form of the box uncertainty set. Here, for each parameter subject to uncertainty, we specify the number of coefficients that can take the worst case deviation. This is unlike the box uncertainty set where every coefficient takes the worst case deviation in an additive manner. For our case, we define the following additional parameters.
    \begin{itemize}
      \item Maximum deviation of inventory holding cost at each CCP: $\delta^{h}_{g}$, $\delta^{h}_{s}$, $\delta^{h}_{r}$, $\delta^{h}_{d}$, $\delta^{h}_{i}$. Units: INR/dose/week.
      \item Maximum deviation of fixed cost of ordering vaccine from a CCP by the next lower-tier CCP: $\delta^{S}_{m}$, $\delta^{S}_{g}$, $\delta^{S}_{s}$, $\delta^{S}_{r}$, $\delta^{S}_{d}$. Units: INR/delivery.
      \item Number of inventory holding cost parameters with uncertainty at each tier: $\tau^{h}_{g}$, $\tau^{h}_{s}$, $\tau^{h}_{r}$, $\tau^{h}_{d}$, $\tau^{h}_{i}$. 
      \item Number of ordering cost parameters with uncertainty at each tier: $\tau^{S}_{m}$, $\tau^{S}_{g}$, $\tau^{S}_{s}$, $\tau^{S}_{r}$, $\tau^{S}_{d}$.
     \end{itemize}
    
    The plain robust version of this problem with budgeted uncertainty set is given below.
    \\
    \textbf{Min $J$} = \\*

    $\mathop{\sum_{t}\sum_{m}\sum_{g}} K^{mg} n_t^{mg} +
    \mathop{\sum_{t}\sum_{g}\sum_{s}} K^{gs} n_t^{gs} +
    \mathop{\sum_{t}\sum_{s}\sum_{r}} K^{sr} n_t^{sr} +
    \mathop{\sum_{t}\sum_{r}\sum_{d}} K^{rd} n_t^{rd} +
    \mathop{\sum_{t}\sum_{d}\sum_{i}} K^{di} n_t^{di} +\\*$
    
    \fbox{Transportation cost $\times$ number of vehicles from one CCP to next lower-tier CCP}\\*
    
    $\mathop{\sum_{t}\sum_{g}} \bar{h}_t^g I_t^g+ \max\limits_{y_{g}^{h}} \delta_{g}^{h} y_{g}^{h} I_t^g +
    \mathop{\sum_{t}\sum_{s}} \bar{h}_t^s I_t^s+
    \max\limits_{y_{s}^{h}} \delta_{s}^{h} y_{s}^{h} I_t^s +
    \mathop{\sum_{t}\sum_{r}} \bar{h}_t^r I_t^r+
    \max\limits_{y_{r}^{h}} \delta_{r}^{h} y_{r}^{h} I_t^r +
    \mathop{\sum_{t}\sum_{d}} \bar{h}_t^d I_t^d+
    \max\limits_{y_{d}^{h}} \delta_{d}^{h} y_{d}^{h} I_t^d +
    \mathop{\sum_{t}\sum_{i}} \bar{h}_t^i I_t^i \; \} +
    \max\limits_{y_{i}^{h}} \delta_{i}^{h} y_{i}^{h} I_t^i +\\*$
    
    \fbox{Holding cost per dose $\times$ inventory at cold chain points}\\

    $\mathop{\sum_{t}\sum_{i}\sum_{j}} p~P^{j}_{t} s^{ij}_{t} +
    \mathop{\sum_{t}\sum_{i}\sum_{j}} \left(1-\eta \right) p~P^j_{t}~ w^{ij}_{t} \\*$
    
    \fbox{Shortage cost per person $\times$ number of persons not vaccinated at clinic }\\
    
    $\mathop{\sum_{t}\sum_{i}\sum_{j}} V^j w^{ij}_{t} +\\*$
    
    \fbox{Clinical services cost per dose $\times$ consumption of vaccine units at clinic }\\
    
    $\mathop{\sum_{t}\sum_{m}\sum_{g}} \bar{S}_t^{mg} x_t^{mg}+ \max\limits_{y_{mg}^{S}} \delta_{mg}^{S} y_{mg}^{S} x_t^{mg} +
    \mathop{\sum_{t}\sum_{g}\sum_{s}} \bar{S}_t^{gs} x_t^{gs}+
    \max\limits_{y_{gs}^{S}} \delta_{gs}^{S} y_{gs}^{S} x_t^{gs} +
    \mathop{\sum_{t}\sum_{s}\sum_{r}} \bar{S}_t^{sr} x_t^{sr}+
    \max\limits_{y_{sr}^{S}} \delta_{sr}^{S} y_{sr}^{S} x_t^{sr} +
    \mathop{\sum_{t}\sum_{r}\sum_{d}} \bar{S}_t^{rd} x_t^{rd}+ 
    \max\limits_{y_{rd}^{S}} \delta_{rd}^{S} y_{rd}^{S} x_t^{rd} +
    \mathop{\sum_{t}\sum_{d}\sum_{i}} \bar{S}_t^{di} x_t^{di} +
    \max\limits_{y_{di}^{S}} \delta_{di}^{S} y_{di}^{S} x_t^{di} +\\*$
    
    \fbox{Ordering cost (cost incurred if even one vaccine unit is ordered)}\\
    
    $\mathop{\sum_{t}\sum_{i}} L n^i_t +
    \mathop{\sum_{t}\sum_{i}} E h^i_t +
    \mathop{\sum_{t}\sum_{i}} F f^i_t$\\
    
    \fbox{Labour costs (wages, hiring and firing costs)}\\
    
    \textbf{Subject to:} \ref{scons1} - \ref{scons10}~ and 
    \begin{equation}\label{eq:21}
        \sum_{a} y_{ab}^{S} = \tau_{a}^{S}~~ \text{if} ~a \in \{m,g,s,r,d\} ~\text{then}~ b \text{ := next lower-tier value from} \{g,s,r,d,i\} 
    \end{equation}
    For example, 
    \begin{equation*}
        \sum_{m} y_{mg}^{S} = \tau_{m}^{S}
    \end{equation*}
    \begin{equation} \label{eq:22} 
        \sum_{a} y_{a}^{h} = \tau_{a}^{h} ~~~a \in \{g,s,r,d,i\}
    \end{equation}
    For example,
    \begin{equation*} 
        \sum_{g} y_{g}^{h} = \tau_{g}^{h} 
    \end{equation*}
    \begin{equation} \label{eq:23}
        0 \leq y_{ab}^{S} \leq 1 ~~~~\forall a ~~~ \text{if} ~a \in \{m,g,s,r,d\} ~\text{then}~ b \text{ := next lower-tier value from} \{g,s,r,d,i\} 
    \end{equation}
    For example,
    \begin{equation*} 
        0 \leq y_{mg}^{S} \leq 1 ~~~~\forall m
    \end{equation*}
    \begin{equation} \label{eq:24}
        0 \leq y_{a}^{h} \leq 1 ~~~~\forall a  ~~~a \in \{g,s,r,d,i\}
    \end{equation}
    For example,
    \begin{equation*} 
        0 \leq y_{g}^{h} \leq 1 ~~~~\forall g
    \end{equation*}
    where $y_{ab}^{S}$ and $y_{a}^{h}$ are binary variables which indicate whether the corresponding coefficients undergo deviation or not. Since the linear relaxation objective function of the inner maximization problem is the same as the original integer program's objective function, we take these variables to be continuous from 0 to 1. We reduce the above bilevel problem to a single level by taking the dual of the inner maximization problem. We now present the single level robust counterpart of the above problem.\\
    
    \textbf{Min $J$} = \\*

    $\mathop{\sum_{t}\sum_{m}\sum_{g}} K^{mg} n_t^{mg} +
    \mathop{\sum_{t}\sum_{g}\sum_{s}} K^{gs} n_t^{gs} +
    \mathop{\sum_{t}\sum_{s}\sum_{r}} K^{sr} n_t^{sr} +
    \mathop{\sum_{t}\sum_{r}\sum_{d}} K^{rd} n_t^{rd}+ \mathop{\sum_{t}\sum_{d}\sum_{i}} K^{di} n_t^{di} +\\*$
    
    \fbox{Transportation cost $\times$ no of trucks from one cold chain point to other}\\*
    
    $\mathop{\sum_{t}\sum_{g} h_t^g I_t^g+ \alpha_{g}^{h} \tau_{g}^{h} + \sum_{g} \beta_{g}^{h}} + \mathop{\sum_{t}\sum_{s} h_t^s I_t^s+ \alpha_{s}^{h} \tau_{s}^{h} + \sum_{s} \beta_{s}^{h}}+ \mathop{\sum_{t}\sum_{r} h_t^r I_t^r+ \alpha_{r}^{h} \tau_{r}^{h} + \sum_{r} \beta_{r}^{h}}+\\*$
    
    $\mathop{\sum_{t}\sum_{d} h_t^d I_t^d+ \alpha_{d}^{h} \tau_{d}^{h} + \sum_{d} \beta_{d}^{h}}+ \mathop{\sum_{t}\sum_{i} h_t^i I_t^i+ \alpha_{i}^{h} \tau_{i}^{h} + \sum_{i} \beta_{i}^{h}}+\\*$
    
    \fbox{Holding cost $\times$ Inventory at cold chain points}\\*
    
    $\mathop{\sum_{t}\sum_{i}\sum_{j}} P^{j}_{t} s^{ij}_{t} +
    \mathop{\sum_{t}\sum_{i}\sum_{j}} \left(1-\eta \right) p~P^j_{t}~ w^{ij}_{t} \\*$
    
    \fbox{Shortage cost $\times$ Shortage of vaccine units at clinic }\\*
    
    $\mathop{\sum_{t}\sum_{i}\sum_{j}} V^j w^{ij}_{t} +\\*$
    
    \fbox{Clinical cost $\times$ Consumption of vaccine units at clinic }\\*
    
    $\mathop{\sum_{t}\sum_{m}\sum_{g} S_t^{mg} x_t^{mg}+ \alpha_{m}^{S} \tau_{m}^{S} + \sum_{m} \beta_{m}^{S}} + \mathop{\sum_{t}\sum_{g}\sum_{s} S_t^{gs} x_t^{gs}+ \alpha_{g}^{S} \tau_{g}^{S} + \sum_{g} \beta_{g}^{S}}+ \mathop{\sum_{t}\sum_{s}\sum_{r} S_t^{sr} x_t^{sr}+ \alpha_{s}^{S} \tau_{s}^{S} + \sum_{s} \beta_{s}^{S}}+\\*$
    
    $\mathop{\sum_{t}\sum_{r}\sum_{d} S_t^{rd} x_t^{rd}+ \alpha_{r}^{S} \tau_{r}^{S} + \sum_{r} \beta_{r}^{S}}+ \mathop{\sum_{t}\sum_{d}\sum_{i} S_t^{di} x_t^{di} + \alpha_{d}^{S} \tau_{d}^{S} + \sum_{d} \beta_{d}^{S}}+\\*$
    
    \fbox{Ordering cost (cost incurred if even one vaccine unit is ordered)}\\*
    
    $\mathop{\sum_{t}\sum_{i}} L n^i_t +
    \mathop{\sum_{t}\sum_{i}} E h^i_t +
    \mathop{\sum_{t}\sum_{i}} F f^i_t$\\
    
    \fbox{Labour costs (Wages, hiring and firing costs)}\\
    
    \textbf{Subject to constraints:} \ref{scons1} - \ref{scons10} ~ and \\
    \begin{equation} \label{eq:25}
        \alpha_{a}^{S} + \beta_{a}^{S} \geq \sum_{b} \sum_{t} x^{ab}_{t}~ \delta_{ab}^{S} ~~~ \forall a  ~ \text{if} ~a \in \{m,g,s,r,d\} ~\text{then}~ b \text{ := next lower-tier value from} ~\{g,s,r,d,i\}
    \end{equation}
    For example,
    \begin{equation*} 
        \alpha_{m}^{S} + \beta_{m}^{S} \geq \sum_{g} \sum_{t} x^{mg}_{t}~ \delta_{mg}^{S} ~~~ \forall m 
    \end{equation*}
    \begin{equation} \label{eq:26}
            \alpha_{a}^{h} + \beta_{a}^{h} \geq \sum_{t} I^{a}_{t}~ \delta_{a}^{h} ~~ \forall a ~~~a \in \{g,s,r,d,i\}
    \end{equation}
    For example, 
    \begin{equation*} 
            \alpha_{g}^{h} + \beta_{g}^{h} \geq \sum_{t} I^{g}_{t}~ \delta_{g}^{h} ~~ \forall g
    \end{equation*}
    \begin{equation} \label{eq:27}
        \beta_{a}^{S} \geq 0 ~~~~~ \forall a ~~~a \in \{m,g,s,r,d\}
    \end{equation}
    For example,
    \begin{equation*} 
        \beta_{m}^{S} \geq 0 ~~~~~ \forall m
    \end{equation*}
    \begin{equation} \label{eq:28}
        \beta_{a}^{h} \geq 0 ~~~~~ \forall a ~~~a \in \{g,s,r,d,i\}
    \end{equation}
    For example,
    \begin{equation*} 
        \beta_{g}^{h} \geq 0 ~~~~~ \forall g
    \end{equation*}
    \begin{equation} \label{eq:29}
        \alpha_{a}^{S} \xrightarrow{} unbounded ~~~a \in \{m,g,s,r,d\}
    \end{equation}
    For example,
    \begin{equation*} 
        \alpha_{m}^{S} \xrightarrow{} unbounded 
    \end{equation*}
    \begin{equation} \label{eq:30}
        \alpha_{a}^{h} \xrightarrow{} unbounded ~~~a \in \{g,s,r,d,i\}
    \end{equation}
    For example,
    \begin{equation*}
        \alpha_{g}^{h} \xrightarrow{} unbounded
    \end{equation*}
    Here $\alpha_{a}^{S}$, $\alpha_{a}^{h}$, $\beta_{a}^{h}$  and $\beta_{a}^{S}$ are the dual variables associated with constraints \ref{eq:25} - \ref{eq:30}.

\section{Model Parameterization: Supplementary Information}
\label{apparam}

Preliminary parameters used in the estimation of optimization formulation parameters (see Table~\ref{table:Final_parameters1} in Section~\ref{subsection:model_parametrization} for model parameter estimates based on these) are provided below. 
\input{Tables/Initial_parameters}

\section{Multiple Vaccine Model: Computational Implementation}
\label{subsection:multi_vaccine_model}
We present the results of the computational implementation (from Section~\ref{section:numerical_results}) of the multiple vaccine model below.
\input{Sections/Multi_commodity_results}

\section{Robust Optimization Analysis for the Single Vaccine Model: Computational Experiments}
\label{aprobust}

The tables providing the results of computational experiments comparing the robust formulation optimal objective function values with the deterministic single vaccine formulation - for the random parameter data instances - are given below (see Section~\ref{subsection:robustresults} for the experiment descriptions). 

\begin{table}[]
\caption{Comparison of objective function values with random parameterization for the deterministic and robust formulations. Notes: w.r.t. = with respect to; min = minimum; avg = average; max = maximum.}
\label{robust-objective-function-value-comparison}
\begin{subtable}{\textwidth}
\caption{Objective function value comparison: low uncertainty.} 
\label{table12a}
\begin{tabular}{|c|c|c|c|c|c|}
\hline
 \multicolumn{1}{|c|}{\multirow{4}{1.5cm}{\textbf{ Data Instance}}} & \multicolumn{3}{|c|}{\textbf{Objective function values ($\times$ 10\textsuperscript{11} INR)}} & \multicolumn{1}{|c|}{\multirow{4}{2.2cm}{\textbf{Savings from box w.r.t. deterministic}}} & \multicolumn{1}{|c|}{\multirow{4}{2.2cm}{\textbf{Savings from box w.r.t. budgeted}}} \\ \cline{2-4}
 & \multicolumn{1}{|c|}{\multirow{3}{1.7cm}{\textbf{Deter- \newline ministic \newline case}}} & \multicolumn{1}{|c|}{\multirow{3}{1.7cm}{\textbf{Robust: box uncertainty}}} & \multicolumn{1}{|c|}{\multirow{3}{2.2cm}{\textbf{Robust: budgeted uncertainty}}} & \multicolumn{1}{|c|}{} & \multicolumn{1}{|c|}{} \\
 & \multicolumn{1}{|c|}{} & \multicolumn{1}{|c|}{} & \multicolumn{1}{|c|}{} & \multicolumn{1}{|c|}{} & \multicolumn{1}{|c|}{}\\ 
 & \multicolumn{1}{|c|}{} & \multicolumn{1}{|c|}{} & \multicolumn{1}{|c|}{} & \multicolumn{1}{|c|}{} & \multicolumn{1}{|c|}{}\\ \hline
 \textbf{Optimal} & 8.7494845 & 8.7495986  & 8.7495895 & -11407257.70 & 903158.80 \\ \hline
\textbf{1} & 8.7494849 & 8.7494793 & 8.7494719 & 563985.81 & 739440.36\\
\textbf{2} & 8.7494847 & 8.7494805 & 8.7494732 & 418337.92 & 731649.16 \\
\textbf{3} & 8.7494857 & 8.7494794 & 8.7494714 & 632768.84 & 800031.42 \\
\textbf{4} & 8.7494847 & 8.7494803 & 8.7494711 & 444728.56  & 916506.33 \\
\textbf{5} & 8.7494852 & 8.7494800 & 8.7494719 & 511858.06 & 812630.18 \\
\textbf{6} & 8.7494858 & 8.7494793 & 8.7494735 & 648470.09 & 582883.34 \\
\textbf{7} & 8.7494854 & 8.7494795 & 8.7494739 & 594246.13 & 562443.31 \\
\textbf{8} & 8.7494851 & 8.7494794 & 8.7494726 & 570654.01 & 672332.75 \\
\textbf{9} & 8.7494842 & 8.7494806 & 8.7494723 & 361341.66 & 824485.18 \\
\textbf{10} & 8.7494853 & 8.7494801 & 8.7494730 & 522370.92 & 715842.28\\ \hline
\textbf{Min} & 8.7494842 & 8.7494793 & 8.7494711 & 361341.65 & 562443.31 \\
\textbf{Avg} & 8.7494851 & 8.7494798 & 8.7494725 & 526876.20 & 735824.43 \\
\textbf{Max} & 8.7494858 & 8.7494806  & 8.7494739 & 648470.09 & 916506.33 \\ \hline                                                                          
\end{tabular}
\end{subtable}
\end{table}

\begin{table}[]
\ContinuedFloat
\label{robust-objective-function-value-comparison}
\begin{subtable}{\textwidth}
\caption{Objective function value comparison: medium uncertainty.} 
\label{table12b}
\begin{tabular}{|c|c|c|c|c|c|}
\hline
 \multicolumn{1}{|c|}{\multirow{4}{1.5cm}{\textbf{Data Instance}}} & \multicolumn{3}{c|}{\textbf{Objective function values ($\times$ 10\textsuperscript{11} INR)}} & \multicolumn{1}{c|}{\multirow{4}{2.2cm}{\textbf{Savings from box w.r.t. deterministic}}} & \multicolumn{1}{c|}{\multirow{4}{2.2cm}{\textbf{Savings from budgeted w.r.t. box}}} \\ \cline{2-4}
 & \multicolumn{1}{c|}{\multirow{3}{1.7cm}{\textbf{Deter- \newline ministic \newline Case}}} & \multicolumn{1}{c|}{\multirow{3}{1.7cm}{\textbf{RO with Box Uncertainty}}} & \multicolumn{1}{c|}{\multirow{3}{2.2cm}{\textbf{RO with Budgeted Uncertainty}}} & \multicolumn{1}{c|}{} & \multicolumn{1}{c|}{} \\
 & \multicolumn{1}{c|}{} & \multicolumn{1}{c|}{} & \multicolumn{1}{c|}{} & \multicolumn{1}{c|}{} & \multicolumn{1}{c|}{}\\ 
 & \multicolumn{1}{c|}{} & \multicolumn{1}{c|}{} & \multicolumn{1}{c|}{} & \multicolumn{1}{c|}{} & \multicolumn{1}{c|}{}\\ \hline
 \textbf{Optimal} & 8.7494845 & 8.7495986  & 8.7495895 & -11407257.70 & 903158.80 \\ \hline
\textbf{1} &	8.7494847 &	8.7494819 &	8.7494740 &	278925.43 &	791754.17 \\
\textbf{2}	& 8.7494851 &	8.7494815 &	8.7494754 &	360520.21 &	614332.42 \\
\textbf{3} &	8.7494850 &	8.7494812 &	8.7494732 &	384622.74 &	796839.63 \\
\textbf{4}	& 8.7494874 &	8.7494795 &	8.7494719 &	796858.27 &	753806.82 \\
\textbf{5} &	8.7494858 &	8.7494822 &	8.7494717 &	362171.20 &	1049977.81 \\
\textbf{6} &	8.7494860 &	8.7494785 &	8.7494704 &	744509.69 &	808824.25 \\
\textbf{7} &	8.7494847 &	8.7494809 &	8.7494727 &	386078.45 &	814890.82 \\
\textbf{8} &	8.7494866 &	8.7494803 &	8.7494706 &	622986.67 &	971418.98 \\
\textbf{9}	& 8.7494850 &	8.7494768 &	8.7494728 &	818478.55 &	397647.07 \\
\textbf{10} &	8.7494871 &	8.7494789 &	8.7494735 &	819194.68 &	538759.51 \\ \hline
\textbf{Min} &	8.7494847 &	8.7494768 &	8.7494704 &	278925.43 &	397647.07 \\
\textbf{Avg} &	8.7494857 &	8.7494802 &	8.7494726 &	557434.59 &	753825.15 \\
\textbf{Max} &	8.7494874 &	8.7494822 &	8.7494754 &	819194.68 &	1049977.81 \\ \hline
                                                                        
\end{tabular}
\end{subtable}
\end{table}

\begin{table}[]
\ContinuedFloat
\label{robust-objective function value comparison in high uncertainty}
\begin{subtable}{\textwidth}
\caption{Objective function value comparison: high uncertainty.} 
\label{table12c}
\begin{tabular}{|c|c|c|c|c|c|}
\hline
 \multicolumn{1}{|c|}{\multirow{4}{1.5cm}{\textbf{ Data Instance}}} & \multicolumn{3}{c|}{\textbf{Objective function values ($\times$ 10\textsuperscript{11}}} & \multicolumn{1}{c|}{\multirow{4}{2.2cm}{\textbf{Savings from box w.r.t. deterministic}}} & \multicolumn{1}{c|}{\multirow{4}{2.2cm}{\textbf{Savings from budgeted w.r.t. box}}} \\ \cline{2-4}
 & \multicolumn{1}{c|}{\multirow{3}{1.7cm}{\textbf{Deter- \newline ministic \newline Case}}} & \multicolumn{1}{c|}{\multirow{3}{1.7cm}{\textbf{RO with Box Uncertainty}}} & \multicolumn{1}{c|}{\multirow{3}{2.2cm}{\textbf{RO with Budgeted Uncertainty}}} & \multicolumn{1}{c|}{} & \multicolumn{1}{c|}{} \\
 & \multicolumn{1}{c|}{} & \multicolumn{1}{c|}{} & \multicolumn{1}{c|}{} & \multicolumn{1}{c|}{} & \multicolumn{1}{c|}{}\\ 
 & \multicolumn{1}{c|}{} & \multicolumn{1}{c|}{} & \multicolumn{1}{c|}{} & \multicolumn{1}{c|}{} & \multicolumn{1}{c|}{}\\ \hline
 \textbf{Optimal} & 8.7494845 & 8.7495986  & 8.7495895 & -11407257.70 & 903158.80 \\ \hline
\textbf{1}	&	8.7494883	&	8.7494797	&	8.7494730	&	857243.55	&	670113.06 \\
\textbf{2}	&	8.7494868	&	8.7494824	&	8.7494742	&	442556.27	&	820460.64 \\
\textbf{3}	&	8.7494879	&	8.7494829	&	8.7494712	&	500555.12	&	1167683.27 \\
\textbf{4}	&	8.7494857	&	8.7494806	&	8.7494685	&	510244.64	&	1209990.55 \\
\textbf{5}	&	8.7494860	&	8.7494782	&	8.7494717	&	781184.27	&	644152.12 \\
\textbf{6}	&	8.7494860	&	8.7494820	&	8.7494739	&	401890.94	&	806757.01 \\
\textbf{7}	&	8.7494864	&	8.7494828	&	8.7494747	&	361491.55	&	812712.97 \\
\textbf{8}	&	8.7494829	&	8.7494804	&	8.7494727	&	245400.37	&	775316.73 \\
\textbf{9}	&	8.7494830	&	8.7494783	&	8.7494683	&	477469.37	&	998984.07 \\
\textbf{1}0	&	8.7494885	&	8.7494797	&	8.7494757	&	879117.37	&	409429.39 \\ \hline
\textbf{Min}	&	8.7494829	&	8.7494782	&	8.7494683	&	245400.37	&	409429.39 \\
\textbf{Avg}	&	8.7494861	&	8.7494807	&	8.7494724	&	545715.35	&	831559.98 \\
\textbf{Max} &	8.7494885	&	8.7494829	&	8.7494757	&	879117.37	&	1209990.55 \\ \hline

\end{tabular}
\end{subtable}
\end{table}

\end{document}

%% file: Sections/Introduction.tex
The COVID-19 pandemic that originated in China rapidly spread throughout the world, and has caused more than five million deaths worldwide \citep{worldometers} and an estimated economic loss of nearly four trillion US dollars \citep{statista}. While moderately effective treatments have been developed, vaccines offer the best chance for a long-term solution to the pandemic. Therefore, a number of vaccines have been successfully developed and vaccination programmes across the world are being operationalized, with multiple large countries having vaccinated more than half their populations \citep{whovaccdash}. Successfully conducting a vaccination programme for a large population entails significant operational challenges \citep{NCCVMRC, azadi2020optimization}. In particular, ensuring efficient distribution of the vaccines from the manufacturer to the medical centers where they are administered to eligible recipients among the public involves logistical challenges across multiple fronts, especially when multiple tiers of the vaccine storage and distribution network need to be traversed. In this context, we present in this paper a decision support tool, based on an integer linear programming framework, that facilitates optimal distribution of vaccines from the manufacturer to the point of administration across a multi-tier vaccine cold chain network. We demonstrate the applicability of this decision support framework for the case of the distribution of COVID-19 vaccines across the state of Bihar in India.

In most countries, vaccines are typically ordered from their manufacturer by a central planning authority such as the central or state government. The vaccines are then routed through one or more storage/distribution facilities before they are delivered to the point of administration, typically a medical center such as a primary health center. Each of these storage/distribution facilities are referred to as a cold chain point, because most vaccines must be stored and transported in refrigerated or sub-zero conditions. For example, in India, the vaccine cold chain of the public health system has multiple tiers: large government medical store depots (GMSDs) maintained by the central government, state level state vaccine stores (SVSs), regional vaccine stores (RVSs), and district vaccine stores (DVSs). The vaccines themselves are typically administered at medical facilities within a district which can be subcenters, primary health centers, community health centers, or district hospitals \citep{NCCVMRC}. Such multi-tier cold chains can be found in many countries. For example, Bangladesh has a four-tier vaccine chain: central vaccine store, district vaccine stores, \textit{Upazila} or subdistrict vaccine stores, with vaccines being administered at the next tier (unions and wards) \citep{guichard2010vaccine}. Niger also has a four-tier vaccine cold chain similar to that of Bangladesh \citep{assi2013removing}. The Indian multi-tier vaccine cold chain is depicted in Figure~\ref{fig:Flow of vaccines in Indian Supply Chain}. Given this, it is evident that a comprehensive decision support framework for optimizing vaccine distribution across multi-tier cold chain networks can prove useful for health planning authorities.

In this paper, we develop such an integer linear programming (ILP) based decision support framework that can optimize, given a planning horizon with discrete time units (e.g., an eight week period) and vaccine demand by recipient subgroup (e.g., of different age groups) for each time unit, the following aspects of vaccine distribution across the cold chain: (a) the cold chain facility from which the next lower-tier cold chain facility must order (if an order is to be placed in a given time unit); (b) quantities of each vaccine (from a set of multiple vaccines available for a given disease) to be ordered at each cold chain tier in each time unit; (c) number of vehicles required to transport the vaccine quantities ordered in each time unit; (d) inventory levels at each cold chain facility in the network; (e) the number of vaccination staff required in each facility in each time unit; and (f) the number of vaccines to be administered to each subgroup in each time unit. The decision support framework attempts to optimize the above decisions by minimizing a total vaccine distribution cost objective function that takes into account fixed and variable costs (where applicable) associated with each of the above decisions. For example, we consider fixed and variable ordering and transportation costs, inventory holding costs, and vaccination workforce resizing costs. We incorporate measures for the priority of each vaccine recipient subgroup by incorporating a shortage cost per recipient in each subgroup: higher the shortage cost, higher the priority of recipients in that subgroup. Vaccine costs and efficacies are also considered, facilitating decisions regarding which vaccine to order for a given disease.

Previous work involving optimization of various aspects of vaccine distribution have focused on three main areas. The first involves optimizing timing of vaccine development (e.g., which strain of the virus to incorporate in the vaccine, as studied by \citet{kornish2008}), and determining vial size and vaccine inventory replenishment schedules so that open vial wastage is minimized \cite{azadi2019developing,azadi2020optimization}. The second area involves optimizing vaccine allocation across multiple recipient subgroups or geographical regions to minimize the impact of a disease \citep{chick2008supply,yarmand2014}. The third area involves optimizing aspects of production and distribution of vaccines. For example, \citet{HOVAV201549} develop a nonlinear integer programming model to optimize distribution of a single vaccine across three tiers. Our work is concerned with the third area, and our research contributions with respect to previous work in this area are detailed in the subsequent section. In addition, we discuss the various approaches adopted in the literature to assess the impact of uncertainty in the vaccine cold chain and their models on the optimal vaccine distribution policies generated by these models. In this context, we discuss our contribution: the development of robust optimization versions of the base model in our decision support framework.

The rest of the paper is organized as follows. In the following section, we present the relevant literature. In Section~\ref{section:methods}, we describe the vaccine distribution problem, and the single-vaccine and multiple vaccine integer programming formulations. We also describe our proposed robust optimization version of the single vaccine model. In Section~\ref{section:numerical_results}, we illustrate the application of the single vaccine and multi-vaccine formulations (for two vaccines) for the COVID-19 case, potential methods to accelerate solution times for the models within the framework, and computational experiments illustrating the performance of the robust version of the single vaccine model when considering multiple uncertainty characterization methods. In Section~\ref{disc}, we conclude the paper.

%% file: Sections/Literature_Review.tex
In this section, we motivate our work with respect to the extant vaccine supply chain management literature. In the vaccine supply chain literature, researchers have focused on the following main research directions: a) vaccine composition, b) allocation of vaccines among the target population, and c) production and distribution of vaccines. We discuss the literature in the first two areas in \ref{aplitreview}, and given the more direct relevance of the third area, we discuss it below. We conclude with an account of the literature on handling parameter uncertainty associated with supply chain models, and a listing of our research contributions with respect to the relevant literature. 


\subsection{Production and Distribution of Vaccines}
Vaccine production is characterised by uncertain production yields and demand, longer manufacturing times and frequent changes in composition of vaccines. \cite{federgruen2008} study a problem in which the decision maker has to satisfy uncertain demand by sourcing vaccines from multiple suppliers who in turn have uncertain production yields. One way to manage uncertainty in production yields and demands is to adjust the pricing and selling strategies. \cite{cho2013} studied three selling strategies namely, advance, regular and dynamic selling. The authors analyse and present the options that are beneficial to the manufacturer and the retailer. \cite{eskandarzadeh2016} extended the work for a risk averse supplier in which the risk of lower production yield is controlled through pricing and quantity as decision variables. With regard to vaccine inventory management at a single site (e.g., a clinic), \cite{lim2017process} develop a lean-inspired set of processes using secondary vaccine packaging and simplified inventory tracking for more efficient inventory management at the site under consideration.

With regard to vaccine distribution across a network of demand sites, \cite{chen2014planning} develop a linear programming model for maximizing the number of fully immunized children. The authors assume that vaccine supply is sufficient to meet demand, consider costs to a very limited extent in their model, and also do not support decision-making regarding transportation or vaccination staff capacity as part of their model. In more recent work, \cite{lim2022redesign} attempt to optimally redesign vaccine distribution networks entirely - that is, they optimize both the number of intermediate distribution centers (aside from the central store and vaccine administration clinics) and their location via a mixed integer formulation. They also determine the number of trips between tiers in their network required to fully meet vaccine demand. The authors motivate their work based on the study by \cite{assi2013removing}, who found via simulation that removing one of the intermediate tiers (the regional tier) of Niger's vaccine cold chain improves the efficiency of vaccine distribution in achieving immunization coverage. 

A key study relevant to our work - which is to develop a comprehensive framework for optimal vaccine distribution across an existing multi-tier vaccine cold chain network - is that by \cite{HOVAV201549}. The authors present a multi-echelon (cost-benefit) model for inventory management of an influenza vaccine supply chain in Israel. The objective of the model is to minimize vaccination costs. The authors present a network flow approach to model the distribution of vaccines across their three-tier supply chain (manufacturers, distribution centers and recipients). A key drawback of their study is that they formulate their problem as a mixed-integer nonlinear optimization problem. Further, they do not consider fixed costs associated with transportation (e.g., the cost of booking a vehicle for transporting vaccines from one cold chain tier to another cold chain tier), or vaccination staffing decisions (they assume a fixed vaccination staff size). Our work addresses these shortcomings by: (a) developing an integer linear program of vaccine distribution across the vaccine cold chain, (b) integrating both fixed and variable transportation and storage costs as well as staffing decisions at vaccination sites within our model, and (c) extending our single vaccine model to consider multiple vaccines for the same condition, which supports decision-making regarding which vaccine to administer to which recipient subgroup. We also modify their conceptualization of shortage costs associated with not receiving a vaccine, which we discuss in Section~\ref{subsection:model_parametrization}. Our work also considers the importance of considering uncertainty in parameters in the overall vaccine supply chain.
\subsection{Uncertainty Modeling in Supply Chain Applications}
The vaccine supply chain is inherently complex, with uncertainties in product selection, production and distribution phases. At the selection stage, uncertainty arises when a public health organization (typically WHO) decides which variant of the virus should be targeted in the current vaccination season. Since vaccine production incurs a long lead time, manufacturers are typically unable to fully satisfy the demand in the vaccination season. The next level of uncertainty occurs in the yield of the production process. The final level of uncertainty occurs in the demand side - for example, vaccine hesitancy in the target population, antigenic drift in the virus, etc. Researchers have primarily studied supply chain contracts as ways to deal with uncertainties arising in these scenarios. \citet{cho2010} studied the impact of yield uncertainty in an influenza vaccine supply chain in an uncertain product scenario - that is, uncertainty associated with which viral strain might cause an outbreak in the current season. The author proposes an optimal dynamic policy to improve social welfare. \citet{cho2013} studied three strategies to mitigate supply side and demand side uncertainty. The first is an advanced selling strategy where selling happens before both supply and demand is realised, the second involves regular selling where selling happens after the demand and supply are realised, and finally, the case of dynamic selling which incorporates both advanced and regular selling strategies. \citet{arifouglu2012} studied the impact of yield uncertainty and demand side uncertainty due to self-interested individuals on an influenza vaccine supply chain. \citet{dai2016} proposes supply chain contracts as a mechanism to address the problem of inefficiencies occurring due to supply side and demand side uncertainties. In recent work, \citet{arifouglu2021} propose vaccination incentives to demand side and a ``menu of transfer payments" to the supply side to mitigate inefficiencies occurring due to uncertainties in the influenza vaccination supply chain. \citet{chandra2021} studied subsidy contracts to coordinate a vaccination supply chain with stochastic production yields. The authors show that subsidy contracts can achieve channel coordination in a vaccination supply chain with varying production yields.

Very few studies have used mathematical optimization techniques to deal with uncertainties arising in vaccination supply chain. \citet{ozaltin2018} studied the composition and production decisions of influenza vaccine as a bilevel multi stage stochastic mixed integer program. At the upper level of the optimization problem, composition decisions (which influenza strains to include in the vaccine) are made while at the second level production decisions are made conditional to the upper level decisions. \citet{sazvar2021} studied a capacity planning problem in designing a resilient supply chain which faces uncertainties due to disruptions. The authors propose a multi-objective optimization model for the same. The authors incorporate redundant capacities as a means to counter risk arising from disruption. In this work, we consider uncertainties in production and distribution by considering uncertainty in multiple parameters associated with vaccine distribution across the cold chain, such as ordering costs, holding costs, demand, production capacities and storage capacities. We adopt a robust optimization approach to model the multi-echelon vaccine supply chain problem. 

In the following subsection, we present the main contributions of our study with respect to the extant literature.

\subsection{Contributions of Our Study}
The main contributions of our work with respect to the literature discussed above are listed below.
    \begin{enumerate}
        \item To the best of our knowledge, our model represents the most comprehensive multi-echelon inventory flow optimization model for a vaccine cold chain in terms of the set of decisions - facility selection at each tier, vaccine choice, vaccine quantities, number of transportation vehicles per time period, vaccination staff, recipient subgroups to be vaccinated, etc. - supported.
        \item Our model is an integer linear program model of the network flow in a vaccine cold chain which, in contrast to the existing nonlinear integer program models developed for similar vaccine cold chain networks (\cite{HOVAV201549}), can yield exact optimal solutions.
        \item Our study is the first inventory flow model for the vaccine cold chain network in the Indian context that incorporates entities at government, state, region, district and clinic levels.
        \item Our model provides a framework for integrating both fixed and variable vaccine transportation costs.
        \item The model allows for determining the optimal vaccination staff levels required to satisfy the vaccine demand in a given time frame.
        \item The model incorporates lead times associated with transportation between cold chain entities as well as the time required to prepare and administer newly arrived batches of vaccines at the clinic tier in the cold chain network.
        \item To the best of our knowledge, our model provides the most comprehensive accounting of vaccine recipient subgroup prioritization considerations to date.
        \item In our knowledge, our framework is the first to provide a robust optimization approach towards modeling uncertainty associated with vaccine distribution across cold chain networks.
    \end{enumerate}
    
Our model can be extended to any vaccine or even multiple vaccines together taking into account the respective capacity requirements of each vaccine. We deal with the tactical decisions of allocation-location as well as strategic decisions of capacity planning, inventory level management, and vaccine recipient subgroup choice. 

%% file: Sections/Problem_description.tex

We develop an integer linear programming based framework for optimizing the decisions that need to be taken with regard to vaccine distribution across a hierarchical cold chain network. We consider decisions associated with the logistics of ordering, transporting, storing, and administering vaccine doses across the cold chain network. In addition to the above set of logistical decisions that are key to any supply chain network, we also introduce vaccination staff capacity planning at the last tier of the cold chain as the number of vaccine units that can be administered at any health centre would depend on the availability of the health workers responsible for doing so. We also consider decisions regarding the prioritization of subgroups of eligible recipients of vaccine units within our framework by associating subgroup-specific unit costs of \textit{not} vaccinating a recipient belonging to each subgroup. We illustrate this by categorizing potential recipients by age. We consider this particular criterion for subgroup formation given its wide use in COVID-19 vaccination policies across the world.

Decision-making in our framework starts from when the vaccines are ready to be transported from the manufacturer(s) through the subsequent tiers at different time periods. We associate one or more costs with every decision that we consider in our model, and hence the models within our framework aim to minimize a stylized total cost of operating a vaccine cold chain subject to cold chain storage capacity, transportation capacity, vaccination staff capacity, and administrative constraints. Given that we illustrate the application of our framework to the cold chain network in India, we now provide a brief overview of the same.

\begin{figure}[h]
        \centering
        \includegraphics[width=4in]{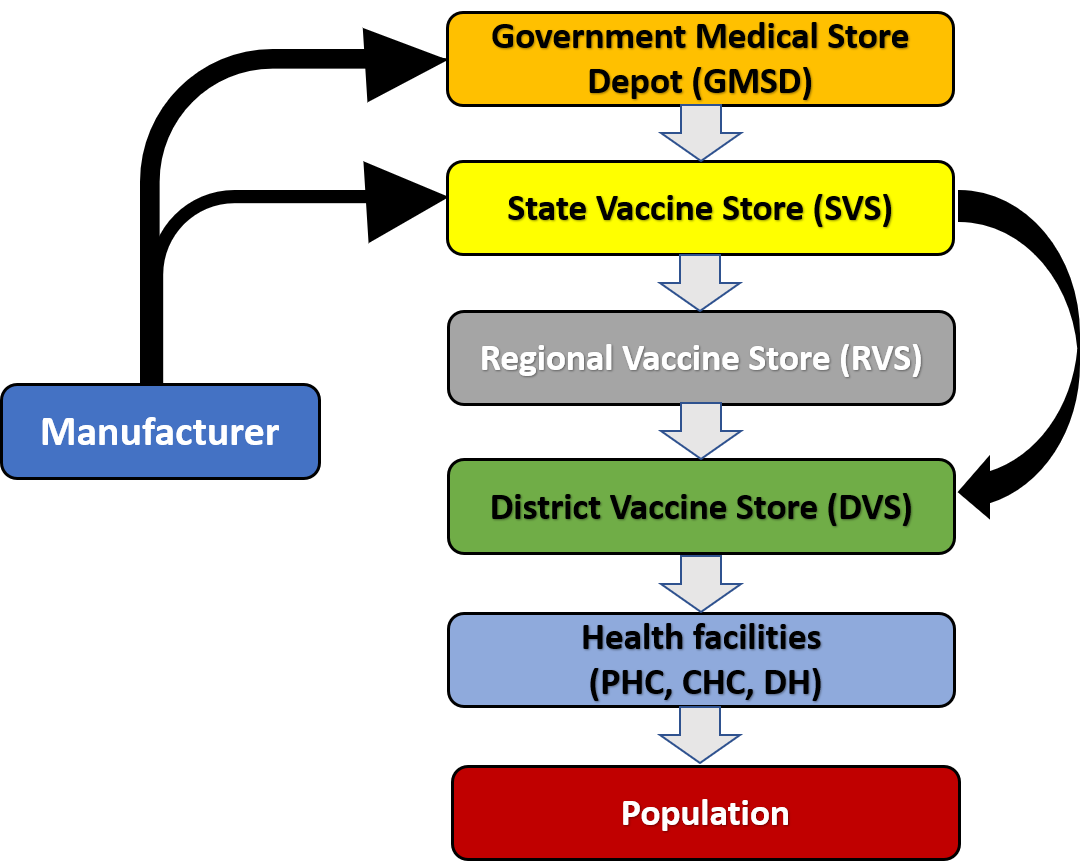}
        \caption{Flow of vaccines in Indian vaccine cold chain. This figure is adapted from the figure on page 3 of the National Cold Chain Assessment India by UNICEF \citep{NCCVMRC}.}
        \label{fig:Flow of vaccines in Indian Supply Chain}
\end{figure}

In India, the vaccine cold chain consists of the following tiers after the manufacturer, listed from the highest tier onwards to the lowest tier: GMSD (Government Medical Store Depot), SVS (State Vaccine Store), RVS (Regional Vaccine Store), DVS (District Vaccine Store) and the clinics (PHCs, CHCs and DHs) where the vaccines are actually administered to the intended recipients. The schematic of the flow of vaccines through the cold chain is shown in Figure \ref{fig:Flow of vaccines in Indian Supply Chain}.

India has 7 GMSDs located mostly in major cities: Mumbai, New Delhi, Chennai, Kolkata, Hyderabad, Guwahati and Karnal. It has 39 state vaccine stores, 123 regional vaccine stores, 644 district vaccine stores, and 20000+ public health centres (which include primary and secondary care facilities) spread across the country \citep{ccps}. For our analysis, we consider the Indian state of Bihar, which has one state vaccine store (SVS) located at its capital city, Patna. We construct a framework that encompasses all the cold chain tiers in Figure~\ref{fig:Flow of vaccines in Indian Supply Chain}; however, our framework can be utilized even if, in practice, one or more tiers do not play a role in a given region in vaccine distribution. From a modeling standpoint, we assume that vaccine manufacturers ship vaccine doses to the GMSDs, which in turn ship to the SVS in the state under consideration. These supply vaccines to the entire state through intermediate levels or tiers comprising of regional vaccine stores (RVS) and district vaccine stores (DVS). DVSs then transport the  vaccines to primary health centres (PHCs) and community health centres (CHCs) which actually administer the vaccine doses to the local population.

In the total operating cost of the cold chain we include vaccine inventory ordering and holding costs at each CCP, fixed and variable transportation costs where fixed costs account for the one-time ordering cost of the vaccine transport vehicles, and the variable costs depend on the distance between CCPs and include fuel, costs of maintaining cold storage in the vehicles, etc. As mentioned earlier, we also include the cost of vaccines, and shortage costs associated with \textit{not} vaccinating an eligible recipient which are different for different population subgroups. We also consider wages, hiring and firing costs of health workers to facilitate vaccination staff capacity planning. We describe how these parameters were estimated later in this section. 

We develop integer linear programs that consider a single vaccine for a given disease as well as multiple vaccines for a disease. We begin by describing the single vaccine formulation. 


%% file: Sections/Mathematical_Formulation_Single_Commodity.tex
We list below all the index sets associated with the single vaccine model, including index sets for each cold chain tier, recipient subgroups and time periods.
 
\begin{itemize}
  \item Manufacturer index, $m \in \{1,2,3,...., M\}$
  \item Government medical store depot index, $g  \in \{1,2,3,...., G\}$
  \item State vaccine store index, $s \in \{1,2,3,...., S\}$
  \item Regional vaccine store index, $r \in \{1,2,3,...., R\} $
  \item District vaccine store index, $d \in \{1,2,3,...., D\}$
  \item Primary health center index, $i \in  \{1,2,3,...., I\} $
  \item Recipient subgroup index, $j \in \{1,2,3,...., J\}$
  \item Time period index, $t \in \{1,2,3,...., T\}$
\end{itemize}

\input{Formulation/parameters_single}
\begin{flushleft}
    \normalsize \textbf{Total Demand}
\end{flushleft}
The following is the total demand by all the subgroups at all the clinics across the entire planning horizon.

$Q$ = $\mathop{\sum_{t}^{T}\sum_{i}^{I}\sum_{j}^{J}}$ $\frac{D^{ij}_{t}}{w}$

\input{Formulation/decision_variables_single}
\input{Formulation/objective_single}
\input{Formulation/constraints_single}

%% file: Formulation/parameters_single.tex
\begin{flushleft}
    \normalsize \textbf{Model parameters}
\end{flushleft}
\begin{itemize}
  \item Inventory holding cost parameters at each cold chain point: $h_t^g$, $h_t^s$, $h_t^r$, $h_t^d$, $h_t^i$. Units: INR/dose/week.
  
  \item Transportation cost (fixed + variable) per vehicle transporting vaccines from one cold chain point to another cold chain point at the next tier (e.g., manufacturer $m$ to GMSD $g$, or DVS $d$ to clinic $i$): $K^{mg}$, $K^{gs}$, $K^{sr}$, $K^{rd}$, $K^{di}$. 
  
  Units: INR/truck. Transportation cost =  (variable costs [diesel costs + labour costs + refrigeration costs] + fixed ordering cost) $\times$ number of vehicles.
    
    \item Vaccine inventory holding capacity at each CCP: $B_t^g$, $B_t^s$, $B_t^r$, $B_t^d$, $B_t^i$. Units: doses.
    \item Capacity of each vehicle transporting  vaccines from one CCP to the next lower-tier CCP (e.g., manufacturer $m$ to GMSD $g$, or DVS $d$ to clinic $i$): $C_t^{mg}$, $C_t^{gs}$, $C_t^{sr}$, $C_t^{rd}$, $C_t^{di}$. Units: doses/vehicle.
    \item Fixed cost (ordering) of ordering vaccine by a CCP at a given level from the next higher-tier CCP (e.g., GMSD $g$ from manufacturer $m$, or clinic $i$ from DVS $d$): $S_t^{mg}$, $S_t^{gs}$, $S_t^{sr}$, $S_t^{rd}$, \textbf{$S_t^{di}$}. Units: INR/delivery.

\item Maximum number of vehicles available for transportation from one CCP to the next lower-tier CCP (e.g., manufacturer $m$ to GMSD $g$, or DVS $d$ to clinic $i$): $N^{mg}$, $N^{gs}$, $N^{sr}$, $N^{rd}$, $N^{di}$. Units: vehicles/week.

    \item Lead times of delivery of vaccines from one CCP to the next lower-tier CCP (e.g., manufacturer $m$ to GMSD $g$, or DVS $d$ to clinic $i$): $L^{mg}$, $L^{gs}$, $L^{sr}$, $L^{rd}$, $L^{di}$. Units: weeks.
    \item Lead times of administration of vaccines at the clinics ($i$): $L^{ij}$. Units: weeks.
  
    \item Initial vaccine inventory held at the CCPs: $I^{g}_{0}$, $I^{s}_{0}$, $I^{r}_{0}$, $I^{d}_{0}$, $I^{i}_{0}$ Units: doses.

  \item Miscellaneous parameters:
  \begin{itemize}
    \item Demand by sub-group $j$ at clinic $i$ at time $t$ (in doses/week):\hfill $D^{ij}_{t}$
    \item Shortage cost (INR) of not vaccinating a customer in subgroup $j$ at time $t$ :\hfill $P^{j}_{t}$
    \item Clinical services cost per customer in subgroup $j$ (e.g., INR/dose):\hfill $V^j$
    \item Average time (e.g., minutes/dose) required for administration of one vaccine dose: \hfill $T_o$
    \item Availability of a health worker in hours at clinic $i$ for time period $t$ (e.g., 40 hours/week):\hfill $N_t^i$
    \item The production capacity of manufacturer $m$ at time $t$ (in doses):\hfill $B_t^m$
    \item Wastage factor (proportion of each dose wasted; that is, effective dose volume required per dose is $\frac{1}{w}$): \hfill $w$
    \item Wages of health workers per time period (e.g., INR/week): \hfill $L$
    \item Fixed cost (INR) of hiring one health worker: \hfill $E$
    \item Fixed cost (INR) of firing one health worker: \hfill $F$
    \item Probability of exposure to the disease-causing pathogen \hfill $p$
    \item Probability of developing the disease after vaccination upon exposure to the pathogen \hfill $\eta$
  \end{itemize}
\end{itemize}

%% file: Formulation/decision_variables_single.tex
\begin{flushleft}
    \normalsize \textbf{Decision Variables}
\end{flushleft}
In the single vaccine formulation, we make decisions regarding the amount of inventory to be held at each CCP across the planning horizon, the number of doses to be ordered by a lower-tier CCP from a higher-tier CCP at each time period, number of vehicles required for transporting the vaccine units from one CCP to a lower-tier CCP, vaccination staff numbers at each clinic, the number of doses to be administered, and the number of people to be vaccinated in each subgroup in each time period.

\begin{itemize}
\item Number of vaccine doses held at each CCP at the end of time $t$: $I^g_t$, $I^s_t$, $I^r_t$, $I^d_t$, $I^i_t$.
\item Number of vaccine doses delivered from one CCP to the next lower-tier CCP at the beginning of time $t$ (e.g., from manufacturer $m$ to GMSD $g$, or from DVS $d$ to clinic $i$): $q^{mg}_t$, $q^{gs}_t$, $q^{sr}_t$, $q^{rd}_t$, $q^{di}_t$.
\item Number of vehicles required for transporting vaccines from one CCP to the next lower-tier CCP  at time $t$ (e.g., manufacturer $m$ to GMSD $g$, or DVS $d$ to clinic $i$): $n_t^{mg}$, $n_t^{gs}$, $n_t^{sr}$, $n_t^{rd}$, $n_t^{di}$.
\item Binary assignment variable indicating whether an order has been placed from a CCP by the next lower-tier CCP (e.g., order placed by GMSD $g$ from manufacturer $m$): $x^{mg}_t$, $x^{gs}_t$, $x^{sr}_t$, $x^{rd}_t$, $x^{di}_t$
\item Number of persons not vaccinated (i.e., number of shortages) and number of vaccine doses administered in subgroup $j$ in clinic $i$ at time $t$, respectively: $s^{ij}_{t}$, $w^{ij}_{t}$
\item Number of health workers working, hired and fired in clinic $i$ at time $t$ respectively: $n^i_t$, $h^i_t$, $f^i_t$

\end{itemize}

%% file: Formulation/objective_single.tex
\begin{flushleft}
    \normalsize \textbf{Objective Function}
\end{flushleft}
The goal of the model is to minimize the total cost associated with vaccine distribution across the cold chain network. The objective function below is constructed from the different subcosts associated with the decisions considered in the cold chain.\\
\textbf{Min $J_{HCO}$} = \\*

$\mathop{\sum_{t}\sum_{m}\sum_{g}} K^{mg} n_t^{mg} +
\mathop{\sum_{t}\sum_{g}\sum_{s}} K^{gs} n_t^{gs} +
\mathop{\sum_{t}\sum_{s}\sum_{r}} K^{sr} n_t^{sr} +
\mathop{\sum_{t}\sum_{r}\sum_{d}} K^{rd} n_t^{rd} +
\mathop{\sum_{t}\sum_{d}\sum_{i}} K^{di} n_t^{di} +\\*$

\fbox{Transportation cost $\times$ number of vehicles from one CCP to next lower-tier CCP}\\*

$\mathop{\sum_{t}\sum_{g}} h_t^g I_t^g+
\mathop{\sum_{t}\sum_{s}} h_t^s I_t^s+
\mathop{\sum_{t}\sum_{r}} h_t^r I_t^r+
\mathop{\sum_{t}\sum_{d}} h_t^d I_t^d+
\mathop{\sum_{t}\sum_{i}} h_t^i I_t^i+\\*$

\fbox{Holding cost per dose $\times$ inventory at cold chain points}\\

$\mathop{\sum_{t}\sum_{i}\sum_{j}} p~P^{j}_{t} s^{ij}_{t} +
\mathop{\sum_{t}\sum_{i}\sum_{j}} \left(1-\eta \right) p~P^j_{t}~ w^{ij}_{t} \\*$

\fbox{Shortage cost per person $\times$ number of persons not vaccinated at clinic }\\

$\mathop{\sum_{t}\sum_{i}\sum_{j}} V^j w^{ij}_{t} +\\*$

\fbox{Clinical services cost per dose $\times$ consumption of vaccine units at clinic }\\

$\mathop{\sum_{t}\sum_{m}\sum_{g}} S_t^{mg} x_t^{mg}+
\mathop{\sum_{t}\sum_{g}\sum_{s}} S_t^{gs} x_t^{gs}+
\mathop{\sum_{t}\sum_{s}\sum_{r}} S_t^{sr} x_t^{sr}+
\mathop{\sum_{t}\sum_{r}\sum_{d}} S_t^{rd} x_t^{rd}+
\mathop{\sum_{t}\sum_{d}\sum_{i}} S_t^{di} x_t^{di} +\\*$

\fbox{Ordering cost (cost incurred if even one vaccine unit is ordered)}\\

$\mathop{\sum_{t}\sum_{i}} L n^i_t +
\mathop{\sum_{t}\sum_{i}} E h^i_t +
\mathop{\sum_{t}\sum_{i}} F f^i_t$\\

\fbox{Labour costs (wages, hiring and firing costs)}

%% file: Formulation/constraints_single.tex
\begin{flushleft}
    \normalsize \textbf{Subject to:}
\end{flushleft}

\begin{itemize}
            
        \item Production capacity constraints associated with the manufacturers. The number of units transported from the manufacturer to the GMSDs cannot exceed the production capacity of the manufacturer at time $t$.
            \begin{equation}
            \label{scons1}
         \mathop{\sum_{g=1}^{G} q^{mg}_t \leq B^m_t} ~~\forall \; m,t
            \end{equation}
            
        \item Facility selection constraints. These constraints restrict the CCP at a given level ($b$) to order only from one CCP at the next higher-tier CCP ($a$).
            \begin{equation}
            \label{scons2}
                \mathop{\sum_{a}} x_t^{ab} \leq 1  ~~ \forall~ b,t ~~~~\text{if } a \in \{g,s,r,d\} ~\text{then}~ b \text{ := next lower-tier value from} \{s,r,d,i\}
            \end{equation}
            For example,
            \begin{equation*}
                \mathop{\sum_{g}} x_t^{gs} \leq 1 ~~ \forall~ s,t 
            \end{equation*}
        
        \item To ensure consistency of $x$ and $q$. These constraints ensure that $x$ and $q$ are zero/non-zero simultaneously.
          \begin{equation}
          \label{scons3}
                q_t^{ab} \leq ( N^{ab} C^{ab}_t )~ x_t^{ab} ~~\forall~ a,b,t ~~~~\text{if } a \in \{m,g,s,r,d\} ~\text{then}~ b \text{ := next lower-tier value from} \{g,s,r,d,i\} ~
            \end{equation}
            For example,
          \begin{equation*}
                q_t^{mg} \leq ( N^{mg} C^{mg}_t )~ x_t^{mg} ~~\forall~ m,g,t
            \end{equation*}
           
           
            
          
        
        \item Transportation vehicle capacity constraints. The number of vehicles needed to transport the requisite number of doses from one CCP to the next lower-tier CCP will depend on the capacity of the vehicles and the number of doses being transported.
            \begin{equation}
            \label{scons4}
                \frac{q_t^{ab}}{C_t^{ab}} \leq  n_t^{ab} ~~\forall~ a,b,t ~~~~\text{if } a \in \{m,g,s,r,d\} ~\text{then}~ b \text{ := next lower-tier value from} \{g,s,r,d,i\}
            \end{equation}
            For example,
            \begin{equation*}
                \frac{q_t^{mg}}{C_t^{mg}} \leq  n_t^{mg} ~~\forall~ m,g,t
            \end{equation*}
            
            
            
        
    
        \item Inventory balance constraints.
        This constraint balances the inflow and outflow of vaccine doses to and from a CCP, taking lead times for vaccine delivery into account. 
            \begin{equation}
            \label{scons5}
              I^b_{t-1} + \mathop{\sum_{a=1}^{A} q^{ab}_{t-L^{ab}} = I^b_t} + \mathop{\sum_{c=1}^{C} q_t^{bc}}~~\forall~ b,t
            \end{equation}
            Here, $\text{if}~b \in \{g,s,r,d\}, \text{then } a :=$ next higher-tier value from $\{m,g,s,r\}$ and $c :=$ next lower-tier value from $\{s, r,d,i\}$.
            For example,
            \begin{equation*}
              I^g_{t-1} + \mathop{\sum_{m=1}^{M} q^{mg}_{t-L^{mg}} = I^g_t} + \mathop{\sum_{s=1}^{S} q_t^{gs}}~~\forall g,t 
            \end{equation*}
            
            
            
            We write the inventory balance constraint for the clinic separately given that vaccine doses are consumed at this CCP.
            \begin{equation*}
               I^i_{t-1} + \mathop{\sum_{d=1}^{D} q^{di}_{t-L^{di}} = I^i_t} + \mathop{\sum_{j=1}^{J} w^{ij}_{t}}~~\forall i,t
            \end{equation*}
         
         \item Inventory Capacity Constraints. The amount of inventory held at a CCP cannot exceed the available capacity of the cold chain equipment at that CCP.
         \begin{equation}
         \label{scons6}
            I^a_t \leq B^a_t ~~\forall~ a,t~~~ a \in \{g,s,r,d,i\}
        \end{equation}
        For example,
        \begin{equation*}
            I^g_t \leq B^g_t ~~\forall~ g,t
        \end{equation*}
           
        \item Consumption balance constraints. The sum of administered doses and the intended doses not administered should be equal to the demand at time $t$.
        \begin{equation}
        \label{scons7}
            \mathop{w^{ij}_{t} + s^{ij}_{t} = \frac{D^{ij}_{t}}{w} } ~~\,\forall~ i,j,t
        \end{equation}
        
        \item Constraints on consumption incorporating lead time of administration. This constraint implies that the number of vaccine doses administered at a clinic $i$ at time $t$ cannot be greater than the inventory level present {$L_i$} periods ago. This constraint implies that a consignment of vaccine doses is likely to be consumed over a certain period of time, and not as soon as the consignment is delivered to the clinic. Note that this constraint is removed if $L_{ij} = 0$.
        \begin{equation}
        \label{scons8}
            \mathop{\sum_{j=1}^{J} w^{ij}_{t} \leq I^i_{t-L^{ij}}} ~~\forall~ i,t
        \end{equation}
        
    \item Medical personnel availability constraints. This constraint restricts the number of vaccine units that can be administered over a time period $t$ based on the availability of vaccination personnel during worker hours and the time it takes to administer a single vaccine dose.
        \begin{equation}
        \label{scons9}
            \mathop{\sum_{j=1}^{J}} T_o w^{ij}_{t} \leq N^{i}_{t} n^{i}_{t} ~~\forall~ i,t
        \end{equation}
        
    \item Health Workers Balance Constraints. A vaccination workforce size balance constraint to model potential movement of vaccination personnel across clinics depending upon the demand at various clinics.
        \begin{equation}
        \label{scons10}
            n^i_t = n^i_{t-1} + h^i_t - f^i_t ~~\forall~ i,t
        \end{equation}

    \end{itemize}

%% file: Sections/robust_methodology.tex
The single vaccine and multiple vaccine formulations that we have presented assume that there is no uncertainty associated with model parameter estimates. However, as outlined earlier, this assumption is unlikely to be realistic in nature. Many model parameter estimates such as ordering costs, holding costs, manufacturing capacities and recipient demand are likely to be subject to uncertainty. Therefore, in order to account for uncertainty in these parameter estimates, we develop a robust optimization version of our single vaccine formulation. 

In our study, we consider two types of uncertainty sets, box and budgeted uncertainty sets, thus yielding two robust formulations. We present the formulations and the additional parameters, decision variables and constraints required in each case. By uncertainty set, we mean that for each concerned parameter, we assume that it can take any value between $[a_{i}- \delta_{i},a_{i}+ \delta_{i}]$, where, $a_{i}$ is the deterministic value and $\delta_{i}$ is the likely deviation of that parameter estimate.

We present the formulation for the box uncertainty set case below, and provide the budgeted uncertainty set formulation in \ref{aprobust}.

\textbf{Box Uncertainty Set Formulation.} 

Box uncertainty set helps us model the worst case scenario. Since the objective of our problem is to minimize the overall cost function, the worst case occurs when the parameters corresponding to the ordering costs and the holding costs at each tier are such that the corresponding objective function terms take the maximum values. The objective function should then be as shown below:
    
    \textbf{Min $J_{HCO}$} = \\*

    $\mathop{\sum_{t}\sum_{m}\sum_{g}} K^{mg} n_t^{mg} +
    \mathop{\sum_{t}\sum_{g}\sum_{s}} K^{gs} n_t^{gs} +
    \mathop{\sum_{t}\sum_{s}\sum_{r}} K^{sr} n_t^{sr} +
    \mathop{\sum_{t}\sum_{r}\sum_{d}} K^{rd} n_t^{rd} +
    \mathop{\sum_{t}\sum_{d}\sum_{i}} K^{di} n_t^{di} +\\*$
    
    \fbox{Transportation cost $\times$ number of vehicles from one CCP to next lower-tier CCP}\\*
    
    $\max\limits_{h_t^{g},h_t^{s},h_t^{r},h_t^{d},h_t^{i}} \{ \;
    \mathop{\sum_{t}\sum_{g}} h_t^g I_t^g+
    \mathop{\sum_{t}\sum_{s}} h_t^s I_t^s+
    \mathop{\sum_{t}\sum_{r}} h_t^r I_t^r+
    \mathop{\sum_{t}\sum_{d}} h_t^d I_t^d+
    \mathop{\sum_{t}\sum_{i}} h_t^i I_t^i \; \} + \\*$
    
    \fbox{Holding cost per dose $\times$ inventory at cold chain points}\\

    $\mathop{\sum_{t}\sum_{i}\sum_{j}} p~P^{j}_{t} s^{ij}_{t} +
    \mathop{\sum_{t}\sum_{i}\sum_{j}} \left(1-\eta \right) p~P^j_{t}~ w^{ij}_{t} \\*$
    
    \fbox{Shortage cost per person $\times$ number of persons not vaccinated at clinic }\\
    
    $\mathop{\sum_{t}\sum_{i}\sum_{j}} V^j w^{ij}_{t} +\\*$
    
    \fbox{Clinical services cost per dose $\times$ consumption of vaccine units at clinic }\\
    
    $\max\limits_{S_t^{mg},S_t^{gs},S_t^{sr},S_t^{rd},S_t^{di}} \{ \;
    \mathop{\sum_{t}\sum_{m}\sum_{g}} S_t^{mg} x_t^{mg}+
    \mathop{\sum_{t}\sum_{g}\sum_{s}} S_t^{gs} x_t^{gs}+
    \mathop{\sum_{t}\sum_{s}\sum_{r}} S_t^{sr} x_t^{sr}+
    \mathop{\sum_{t}\sum_{r}\sum_{d}} S_t^{rd} x_t^{rd}+ \\
    \mathop{\sum_{t}\sum_{d}\sum_{i}} S_t^{di} x_t^{di} \; \} +\\*$
    
    \fbox{Ordering cost (cost incurred if even one vaccine unit is ordered)}\\
    
    $\mathop{\sum_{t}\sum_{i}} L n^i_t +
    \mathop{\sum_{t}\sum_{i}} E h^i_t +
    \mathop{\sum_{t}\sum_{i}} F f^i_t$\\
    
    \fbox{Labour costs (wages, hiring and firing costs)}\\
    
In the above formulation, note that $\bar{S}^{mg}_t$, $\bar{S}^{mg}_t$, $\bar{S}^{gs}_t$, $\bar{S}^{sr}_t$, $\bar{S}^{rd}_t$, $\bar{S}^{di}_t$, $\bar{h}^g_t$, $\bar{h}^s_t$, $\bar{h}^r_t$, $\bar{h}^d_t$ and $\bar{h}^i_t$ are the deterministic values and $\delta_{mg}^{S}$, $\delta_{gs}^{S}$, $\delta_{sr}^{S}$, $\delta_{rd}^{S}$, $\delta_{di}^{S}$, $\delta_{g}^{h}$, $\delta_{s}^{h}$, $\delta_{r}^{h}$, $\delta_{d}^{h}$ and $\delta_{i}^{h}$ are the maximum deviations of the respective parameters. 

Further, one can easily note from the above formulation that the objective function will achieve its worst case (maximum in this case) value when worst case deviations are allowed for all the coefficients. Therefore, the robust counterpart for box uncertainty case is as shown below.\\
    
\textbf{Min $J$} = \\*

    $\mathop{\sum_{t}\sum_{m}\sum_{g}} K^{mg} n_t^{mg} +
    \mathop{\sum_{t}\sum_{g}\sum_{s}} K^{gs} n_t^{gs} +
    \mathop{\sum_{t}\sum_{s}\sum_{r}} K^{sr} n_t^{sr} +
    \mathop{\sum_{t}\sum_{r}\sum_{d}} K^{rd} n_t^{rd} +
    \mathop{\sum_{t}\sum_{d}\sum_{i}} K^{di} n_t^{di} +\\*$
    
    \fbox{Transportation cost $\times$ number of vehicles from one CCP to next lower-tier CCP}\\*
    
    $\mathop{\sum_{t}\sum_{g}} (\bar{h}_t^g+\delta_{g}^{h}) I_t^g+
    \mathop{\sum_{t}\sum_{s}} (\bar{h}_t^s+\delta_{s}^{h}) I_t^s+
    \mathop{\sum_{t}\sum_{r}} (\bar{h}_t^r+\delta_{r}^{h}) I_t^r+
    \mathop{\sum_{t}\sum_{d}} (\bar{h}_t^d+\delta_{d}^{h}) I_t^d+
    \mathop{\sum_{t}\sum_{i}} (\bar{h}_t^i+\delta_{i}^{h}) I_t^i+\\*$
    
    \fbox{Holding cost per dose $\times$ inventory at cold chain points}\\
    
    $\mathop{\sum_{t}\sum_{i}\sum_{j}} p~P^{j}_{t} s^{ij}_{t} +
    \mathop{\sum_{t}\sum_{i}\sum_{j}} \left(1-\eta \right) p~P^j_{t}~ w^{ij}_{t} \\*$
    
    \fbox{Shortage cost per person $\times$ number of persons not vaccinated at clinic }\\
    
    $\mathop{\sum_{t}\sum_{i}\sum_{j}} V^j w^{ij}_{t} +\\*$
    
    \fbox{Clinical services cost per dose $\times$ consumption of vaccine units at clinic }\\
    
    $\mathop{\sum_{t}\sum_{m}\sum_{g}} (\bar{S}_t^{mg}+\delta_{mg}^{S}) x_t^{mg}+
    \mathop{\sum_{t}\sum_{g}\sum_{s}} (\bar{S}_t^{gs}+\delta_{gs}^{S}) x_t^{gs}+
    \mathop{\sum_{t}\sum_{s}\sum_{r}} (\bar{S}_t^{sr}+\delta_{sr}^{S}) x_t^{sr}+
    \mathop{\sum_{t}\sum_{r}\sum_{d}} (\bar{S}_t^{rd}+\delta_{rd}^{S}) x_t^{rd}+
    \mathop{\sum_{t}\sum_{d}\sum_{i}} (\bar{S}_t^{di}+\delta_{di}^{S}) x_t^{di} +\\*$
    
    \fbox{Ordering cost (cost incurred if even one vaccine unit is ordered)}\\
    
    $\mathop{\sum_{t}\sum_{i}} L n^i_t +
    \mathop{\sum_{t}\sum_{i}} E h^i_t +
    \mathop{\sum_{t}\sum_{i}} F f^i_t$\\
    
    \fbox{Labour costs (wages, hiring and firing costs)}\\
    
    \textbf{Subject to constraints:} \ref{scons1} - \ref{scons10}. \\
    
We now describe how the parameters associated with our formulations are estimated.

%% file: Sections/Estimation_of_parameters.tex
\begin{figure}[h]
    \centering
    \includegraphics[width=5in]{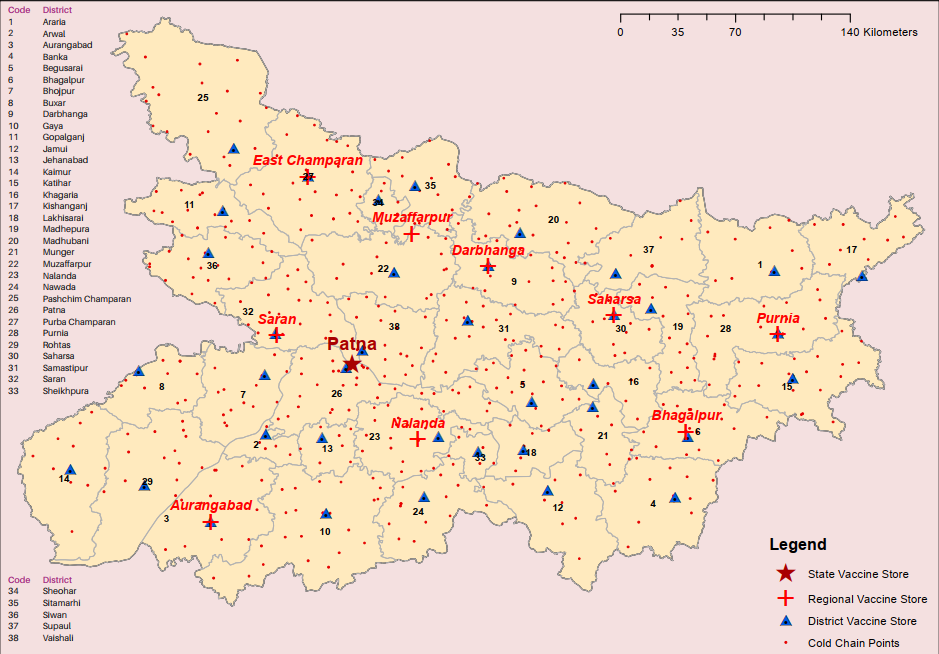}
    \caption{Map of vaccine cold chain network in Bihar, This figure is adapted from the Immunization Cold Chain and Vaccine Logistics Network Factbook \citep{NHM}.}
    \label{fig:Map}
\end{figure}

To illustrate the implementation of our framework, we work witht the Indian state of Bihar. Thus the vaccine cold chain network and associated equipment data we have used to parameterize the models within our framework is that associated with Bihar. The vaccine delivery network in Bihar operates through 1 SVS, 9 RVSs, 38 DVSs, and 606 clinics \citep{NHM}. The map of the cold chain network in Bihar is shown in Figure \ref{fig:Map}. 
\begin{flushleft}
    \normalsize \textbf{Estimation of cold chain and transportation vehicle capacity}
\end{flushleft}
Thus, there are 654 cold chain points within the vaccine distribution network in the state of Bihar. We determined that a total of 1,946 cold chain equipment - such as walk-in freezers, coolers, and ice-lined refrigerators - are used across the cold chain network in the state \citep{NHM}. The numbers of each equipment type, the dimensions of the refrigerated and insulated vans \citep{NCCVMRC} and the cold boxes \citep{NHM-UNICEF} and estimates of other associated parameters such as the utilization factor are given in Table \ref{table:Initial_parameters1} in \ref{apparam}. We have assumed that the cold chain equipment for a given district is uniformly distributed across all its health centers where the vaccines are administered. For calculating the capacity of these cold chain equipment, the packed vaccine volumes per dose for the two types of COVID-19 vaccines were taken from official data released by the Indian government \citep{vaccineindia}.

Vaccines are transported in refrigerated vans from the manufacturer to the GMSD and from the GMSD to the SVS (for long distances), and in insulated vans from the SVS to all other downstream cold chain points \citep{NCCVMRC}. The capacities of these vans were calculated by assuming standard dimensions of the vehicles with a certain utilization factor. Utilization factor is a number less than 1 which is multiplied with the storage shelf volume to arrive at the actual fraction of space available for storing vaccines. It is based on the fact that the entire storage space available for storing vaccines cannot be used due to loss of vaccine doses caused by vaccine handling practices, packaging dimensions, etc. The most commonly used estimate for this parameter is 0.67 \citep{world2017calculate}. The distances between each CCP were estimated via the Bing Maps application, and used to populate a distance matrix. The distance is then multiplied by fuel (diesel) cost to get the variable transportation cost. Fixed transportation costs have been reasonably assumed to account for the one-time ordering cost of a vehicle.
\begin{flushleft}
    \normalsize \textbf{Estimation of demand}
\end{flushleft}

In order to demonstrate how the optimization framework can be used to prioritize certain subpopulations over others, we assumed that the population of the state can be divided into three subgroups categorized on the basis of age: children (less than 18 years), adults (between 18 to 60 years) and elderly (60 years and above). Assuming one dose of vaccine administered per person within a given planning horizon, the weekly demand for each subgroup at each clinic is calculated from the population distribution by age among all the districts in the state. The population distribution by age was obtained from the most recent census data published by the Indian government \citep{censusdata}. This is multiplied with the growth rates by age group \citep{growthrate} to arrive at the estimates of the population size of each subgroup.
\begin{flushleft}
    \normalsize \textbf{Estimation of other parameters}
\end{flushleft}

The production capacity of the manufacturer has been assumed to be around 2 billion doses for the entire country \citep{europeanpharmaceuticals}. However, as shown in Section~\ref{subsection:single_vaccine_model}, we analyze how the manufacturing capacity affects the extent to which demand is satisfied across the planning horizon.

The inventory holding cost has been assumed to be Rs 0.3 per unit vaccine per week for all CCPs at all the tiers. The ordering costs at each tier have been reasonably assumed with costs per order increasing with the level of the CCP tier. We note here that the vaccine ordering costs at each cold chain tier can be considered to be proxy measures of the efficiency of the ordering process at a given tier. Thus these costs can be adjusted depending upon the perception of the decision-maker or analyst using this proposed decision support tool regarding the efficiency of the ordering process at a given cold chain tier. Note that ordering costs can be varied across specific facilities within a cold chain tier. Hence it is the relative values of the ordering costs both within and across cold chain tiers that are of more importance than the actual estimates themselves. A similar argument can be made for the fixed costs associated with transportation as well. 

The wages, hiring, and firing cost of vaccination staff have been reasonably assumed by collecting data on standard wages. The wages are taken as the median salary of nurses (monthly) in Bihar which are adjusted appropriately to obtain the weekly wages. The hiring and firing costs have been accordingly assumed as some proportion of the monthly salary of the nurses.

We explain the estimation of the shortage costs within the context for the multiple vaccine model, which we can consider as a generalization of the single vaccine model. We develop our modeling of the costs of not vaccinating eligible recipients based on the notion of shortage costs introduced in \cite{HOVAV201549}; however, we extend their conceptualization in the following ways: (a) we modify their shortage cost model to incorporate the multiple vaccine case, which involves including a measure of vaccine effectiveness in the shortage cost model; and (b) we include the costs of the loss of life due to the disease in question, as well as the costs of illness (but not mortality), among both vaccinated and unvaccinated persons. The shortage costs are included in the objective function of the multi-vaccine formulation in the following manner:
\begin{equation*}
 \mathop{\sum_{t}\sum_{i}\sum_{j}} p~P^j_{t} s^{ij}_{t} + \mathop{\sum_{t}\sum_{k}\sum_{i}\sum_{j}} p~ \left(1-\eta_{k}\right)~P^j_{t}~ w^{ijk}_{t}\\*
\end{equation*}
Here $p$ is the probability of exposure to the SARS-CoV2 virus (we assume exposure to the virus leads to developing symptomatic or asymptomatic COVID-19 with 100\% probability), and $\eta_{k}$ is the effectiveness of the vaccine in preventing symptomatic infection. Thus $1 - \eta_{k}$ represents the probability of developing COVID-19 upon exposure to the virus after getting vaccinated. The probability of exposure to the disease $p$ can be estimated using serosurvey data (for example, 56\% of the population residing in certain areas of New Delhi were found in a serosurvey to have COVID-19 antibodies \citep{serosurvey1} or can even be set to 1.0 for endemic diseases. We multiply both terms by $P^j_t$ (the shortage cost associated with a person in the $j^{th}$ subgroup developing a symptomatic form of the disease under question) to get the costs incurred in each case. $P^j_t$ is given by the following formula:
\begin{equation*}
    P^{j}_{t} = \left[ m I \left(L - \Bar{A} \right) + \left(1-m\right)*R\right]
\end{equation*}
Here $m$ is the case fatality rate for the disease, $I$ is the per-capita income, $\Bar{A}$ is the average age in the $j^{th}$ subgroup, $L$ is the population average life expectancy and $R$ is the cost of treatment incurred by the payer (e.g., the government, or the societal cost as a whole) per case of the disease. Vaccine effectiveness ($\eta_k$) values have been taken from the official data given for each vaccine. A sero-survey (which tested people for COVID-19 antibodies via serological tests) was conducted in Delhi in the month of January 2021, and estimated that about 56\% of the Delhi residents were exposed to the SARS-CoV2 virus \citep{serosurvey1}. This gives us an estimate of $p$, the probability that a person is exposed to the virus. We considered the most recent estimate of the case fatality rate of COVID-19, found out the difference between the average age of the subgroup and life expectancy ($\sim$ 69 years)~\citep{data} and multiplied this with the per capita income (Rs 1,25,408)~\citep{percapita} of India to estimate the shortage cost in case the person dies from COVID-19. If the person does not die (the probability of survival is $1 - m$), then additional treatment fees are incurred which is given by $R$. These estimates were rounded to obtain the shortage costs actually used in the model. We would like to emphasize here that we provide these details to illustrate how shortage costs for a particular vaccine can be calculated and that this may be modified based on the disease under consideration as well as the data available for the vaccine(s) and the disease. For example, we assume that the mortality probability remains the same for vaccinated and unvaccinated people; however, this can be modified easily if required.

The fraction of available CCP and transportation vehicle storage capacity for COVID-19 vaccine has been estimated as a ratio of the total demand for COVID-19 to the total demand for all the other vaccines, taking into account their respective packed vaccine volumes per dose. We note here that a single average packed vaccine volume per dose is taken for all the other vaccines for ease of calculation. 

The values for each of the estimated/assumed parameters are given in Table ~\ref{table:Final_parameters1}.
\input{Tables/Final_values}

We now discuss the computational implementation of the model framework, example analyses that can be carried out using the single vaccine and multiple vaccine models, and preprocessing methods to decrease the computational overhead of the single vaccine model.

%% file: Tables/Final_values.tex
\begin{table}[]
\centering
\caption{Optimization model parameter estimates. \textsuperscript{a}Assumed. \textsuperscript{e}Estimated. \textsuperscript{1}The values indicated here are the demand values of each subgroup averaged over all the districts of Bihar}
\label{table:Final_parameters1}
\footnotesize
\begin{tabular}{|l|c|c|}
\hline
\multicolumn{2}{|c|}{Parameter}  & Estimate \\ \hline
Inventory holding costs (INR/dose/week)\textsuperscript{a}  &  For all CCPs    & 0.3 \\ \hline
\multirow{5}{*}{Fixed Transportation cost (INR)\textsuperscript{a}} & $m \to g$  & 40,000 \\ 
                                                 & $g \to s$ & 20,000 \\ 
                                                 & $s \to r$  & 12,000 \\ 
                                                 & $r \to d$ & 10,000 \\ 
                                                 & $d \to i$ & 5,000  \\ \hline
Diesel cost (INR/km) \textsuperscript{e} &   Across the cold chain               & 14               \\ \hline
\multirow{3}{*}{Packed vaccine volume (cm3/dose) \textsuperscript{e}}   & Covishield & 0.2109  \\
& Covaxin & 0.086 \\
    & Measles & 3.3 \\ \hline
\multirow{5}{*}{Ordering cost (INR/delivery)\textsuperscript{a}}  & $m \to g$ & 200,000  \\
                                               & $g \to s$ & 100,000   \\
                                               & $s \to r$ & 75,000    \\
                                               & $r \to d$  & 25,000    \\
                                               & $d \to i$ & 15,000  \\ \hline
\multirow{3}{*}{Demand $D_t^{ij}$ (doses/week)\textsuperscript{e} \textsuperscript{1}}  & Children (\textless{}18)   & $\sim$ 100,000  \\
                                 & Adults (18 - 60) & $\sim$ 100,000 \\
            & Elderly (\textgreater{}60) & $\sim$ 15,000  \\ \hline
\multirow{3}{*}{Shortage cost $P_t^{j}$
(INR) \textsuperscript{e} }     & Children (\textless{}18)   & 216,661               \\
                                & Adults (18 - 60) & 285,814  \\
        & Elderly (\textgreater{}60) & 322,729  \\ \hline
\multirow{2}{*}{Transportation capacities ($cm^3$/truck) \textsuperscript{e}} & Insulated van              & 960,000  \\
    & Refrigerated van           & 9,675,850               \\ \hline
\multirow{2}{*}{Cost of vaccine (INR/dose) \textsuperscript{e}} & Covishield  & 780   \\
    &    Covaxin                        & 1,410          \\  \hline
\multirow{2}{*}{Effectiveness of vaccine $\eta_k$ \textsuperscript{e}} & Covishield  & 0.937   \\
    &    Covaxin                        & 0.778          \\  \hline
\multicolumn{2}{|l|}{Probability of getting exposed to SARS-CoV2 $p$  \textsuperscript{e}}    & 0.56    \\ \hline  
\multicolumn{2}{|l|}{Time required for administering a dose of vaccine (min/dose)  \textsuperscript{a} }  & 5  \\ \hline
\multicolumn{2}{|l|}{ Availability of medical personnel administering vaccines   (min/week)   \textsuperscript{a}}  & 3,360   \\ \hline
\multicolumn{2}{|l|}{Cost of hiring of medical personnel administering vaccines (INR) \textsuperscript{a}       }  & 5,000  \\ \hline
\multicolumn{2}{|l|}{Cost of firing of medical personnel administering vaccines (INR) \textsuperscript{a}       }  & 2,000  \\ \hline
\multicolumn{2}{|l|}{Weekly wages of medical personnel administering vaccines (INR) \textsuperscript{e}       }  & 6,175  \\ \hline
\multicolumn{2}{|l|}{Vaccine production capacity of the manufacturers (doses/year) \textsuperscript{e}  }  & 2,000,000,000  \\ \hline
\multicolumn{2}{|l|}{Fraction of capacity reserved for storing COVID-19 vaccines  \textsuperscript{e}}    & 0.59    \\ \hline               
\end{tabular}
\end{table}

%% file: Sections/Single_commodity_results.tex
We begin by demonstrating the output of the single vaccine model within the decision support framework that we develop. In order to illustrate how the output of the decision support framework can be organized and analyzed, we consider a relatively small component of the cold chain in the state of Bihar, especially at the district level: we include only two districts (which we refer to henceforth as districts 1 and 2, respectively). This implies that in addition to the manufacturer, the GMSD, the SVS, and 9 RVSs, we consider 2 DVSs and 16 clinics, with 10 located in district 1 and 6 in district 2. Further, for ease of representation of the model output, we consider a 6 week planning horizon. 

We present the output of the single vaccine model for the above cold chain for two cases: in the first case, all model parameters are estimated from Table \ref{table:Final_parameters1} (which we refer to as the base case), and in the second case, we consider a vaccine with a higher packed volume per dose of 3.3 $cm^3$/dose (for the measles vaccine). We present the model output in this manner because the COVID-19 vaccine packed volumes per dose (e.g., 0.211 $cm^3$/dose) appear to be significantly lower than those of other commonly used vaccines, implying that the capacity required to transport and store vaccines other than the COVID-19 vaccines is likely to be significantly lower. This also provides us with an opportunity to demonstrate how storage and transportation capacity affects the optimal ordering, inventory storage, shortages, and staffing decisions associated with vaccine distribution across the cold chain. The output for the base case is provided in Table~\ref{table:single-commodity-lower-packed-volume} and the output for the higher packed volume per dose case is provided in Table~\ref{table:single-commodity-higher-packed-volume}. Tables \ref{table:single-commodity-lower-packed-volume} and  \ref{table:single-commodity-higher-packed-volume} do not list the decisions for every facility in the cold chain network; from the sake of brevity, they only contain the facilities between which vaccines are transported in a given time period. For example, for the base case, we see that all of the ordering and transportation occurs in week 1, and that the vaccines are transported to the clinics along the following path: manufacturer $\to$ the GMSD $\to$ the SVS $\to$ RVS 5 $\to$ DVS $\to$ the clinics.

We first note from Tables \ref{table:single-commodity-lower-packed-volume} and  \ref{table:single-commodity-higher-packed-volume} that the number of shortages incurred (Tables \ref{table:single-commodity-lower-packed-volume-shortage},\ref{table:single-commodity-higher-packed-volume-shortage}), the inventory held at different time periods across the planning horizon at different CCPs, which we refer to henceforth as the inventory pattern (Tables \ref{table:single-commodity-lower-packed-volume-inventory}, \ref{table:single-commodity-higher-packed-volume-inventory}) and the vaccination staff's recruitment schedule (tables \ref{table:single-commodity-lower-packed-volume-staff-planning}, \ref{table:single-commodity-higher-packed-volume-staff-planning}) remain the same for both the cases. However, we see that the ordering and vaccine transportation patterns as seen in Tables \ref{table:single-commodity-lower-packed-volume-ordering} and \ref{table:single-commodity-higher-packed-volume-ordering} are different when the storage and transportation capacity are significantly different. 

We observe that for the base case, a single DVS (DVS 1) supplies the vaccine units to all the 16 clinics. It receives the entire supply from a single RVS (RVS 5), which happens to be the nearest RVS to the SVS. The reason for a single cold chain point handling the entire supply in the RVS and DVS tiers is the higher capacity available for transportation due to lower packed volumes (more than 2.5 million doses per vehicle from both the district and the regional level). All the 16 clinics order from only district 1 because the shortest route, in terms of the sum of the distances of the clinics and RVSs from DVS 1 and DVS 2, is the lowest for RVS 5 and DVS 1 among all possible routes.

However, this does not hold for the case with the higher packed volume per dose. A single district cannot cater to the demand of all 16 clinics because of the reduced transportation capacity available for vaccines with higher packed volume per dose. A vehicle from a DVS can only transport 171,736 doses at a time, which is less than the combined demand of all the clinics (333,616 doses). Therefore, the supply gets split among the two districts, with both districts ordering their respective vaccine quantities from RVSs 5 and 6, which are the nearest RVSs to the SVS. Also, we notice that `cross-ordering' from the DVSs occurs at the clinic level, which means that certain clinics receive vaccines from DVSs in districts other than the district in which they are located. This happens primarily because of the restriction on transportation capacity. If the clinics were to order from the DVSs in their district only, then additional orders would need to be placed, which in turn result in additional ordering and transportation costs. This does occur given the overall cost minimization objective across the cold chain. This analysis thus illustrates the interplay of the ordering costs, transportation distances, and vaccine transport vehicle capacity. Therefore, the optimal ordering and vaccine transport pattern that our proposed framework yields does not necessarily conform to the `shortest' paths (based on inter-facility distances) across the cold chain. Other factors, an example being the vaccine transport vehicle capacities, also play an important role in guiding the logistics of distribution and administration of vaccines.


We also notice from tables \ref{table:single-commodity-lower-packed-volume-shortage} and \ref{table:single-commodity-higher-packed-volume-shortage} that there are shortages in weeks 1 and 2 in both the cases, which result in shortage costs being incurred. This is seen because we have assumed a vaccine delivery lead time of 1 week from DVSs to clinics, and we also assume that one week is required to prepare a newly arrived batch of vaccines at the clinic level so that it is ready for administration to the set of eligible recipients served at that clinic. Therefore, there is a delay of two weeks before the first set of doses get administered. In order to avoid this delay and the subsequent shortage costs incurred, an analyst using this model can simply set the planning horizon to begin the required number of time periods (depending upon the lead times associated with vaccine delivery from one tier to the next lower tier) before the actual demand is incurred, and can set the demand during this `lead time' period to be zero. We also note that our current assumption of lead times at only the DVS and clinic level is only to illustrate the impact of lead times on the shortage patterns across the planning horizon; depending upon the vaccine ordering, transportation and vaccine administration patterns, lead times may need to be incorporated at other tiers also.
\input{Tables/single_commodity_COVID}
\input{Tables/single_commodity_measles}

In continuation with the above analyses, we study the impact of cold chain capacity in more detail - storage and transportation vehicle capacity - on the optimal shortage and inventory patterns generated by our model. For this, using the high packed volume per dose case (as packed volumes per dose higher than that of the COVID-19 vaccines appear to be more prevalent), we vary the fraction of the storage and transportation capacity available for the vaccine under consideration and report the total shortages incurred at the end of the planning horizon and the inventory pattern at each cold chain tier. We find that for a given parameterization of the model, there exists a threshold or critical value of this fraction (0.15 in our case) above which the number of shortages incurred becomes constant. Below this fraction, the number of shortages increases and consequently result in increasing shortage costs as well. In addition to the decreased storage capacity, the cap on the number of vehicles that are available at a cold chain point for transportation to the next lower-tier cold chain point also leads to this increase in the number of shortages incurred. Further, the critical fraction of cold chain capacity below which the shortages start increasing remains the same regardless of the lead times assumed across the cold chain, as this increase in the shortages is only due to the limits on storage and transportation capacity and is not related to the lead time.

We also note that even at low capacities, the proportion of eligible recipients in each subgroup not receiving the vaccines is the lowest in the highest priority subgroup (close to 0\% in elderly) and the highest in the lowest priority subgroup (close to 100\%) at lower fractions of available capacity. In other words, the subgroup with a higher shortage cost (a proxy for higher priority) is always catered to first, after which other subgroups are considered. Note that an optimization formulation is not necessarily required to determine which subgroups should be vaccinated first once the shortage costs are determined, at least for the single vaccine. However, in situations where there are multiple vaccines for a single disease (i.e., the multiple vaccine model presented in this paper), the shortage costs become relevant, as we discuss in the following section, to determine which vaccine is to be administered, especially if there are trade-offs between vaccine efficacy and various associated costs - the cost per dose of the vaccine itself, its holding cost, ordering cost, transportation costs, etc. The interplay between the shortage costs, efficacy, and all the associated costs listed above may prove difficult to unravel without a formulation such as that we present here. Further, we included it as a placeholder for the case where the single vaccine formulation is extended to consider multiple vaccines for multiple diseases. In this case, similar or overlapping subgroups on the basis of demographic characteristics might be present, implying that using shortage costs in conjunction with vaccine efficacies and their costs might become necessary to determine the optimal set of vaccines and the associated subgroups to vaccinate.

We also briefly discuss the impact of this parameter on the inventory pattern across the cold chain. At lower values of this parameter, since the shortages are very high, the number of vaccines ordered at each tier is very less, which subsequently results in lower inventory levels. At higher values, since the capacity of vehicles is now higher, given the fact that more numbers of vaccines can be transported in fewer orders due to the high capacities, the algorithm tries to transport all the vaccines further to the next level as soon as it receives the order leading to zero inventory levels at all tiers. As we decrease the value of this parameter from the higher end, due to the reduced vaccine transportation capacities across tiers, it gets stored in the inventory and hence inventory level increases. Higher levels of inventory are seen for the GMSD and the SVS compared to the other CCPs due to the higher ordering costs at these cold chain tiers.

\begin{figure}
     \centering
     \begin{subfigure}[b]{0.4\textwidth}
         \centering
         \includegraphics[width=\textwidth]{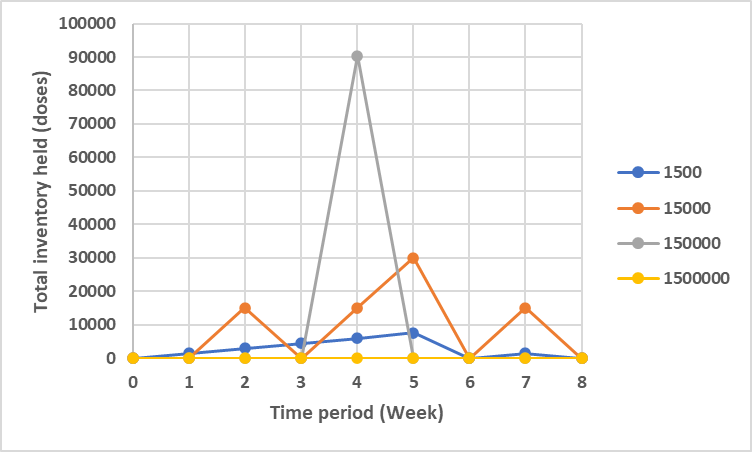}
         \caption{Vaccine inventory held at the GMSD.}
         \label{fig:Total inventory held at GMSD in doses across the planning horizon}
     \end{subfigure}
     \hfill
     \begin{subfigure}[b]{0.4\textwidth}
         \centering
         \includegraphics[width=\textwidth]{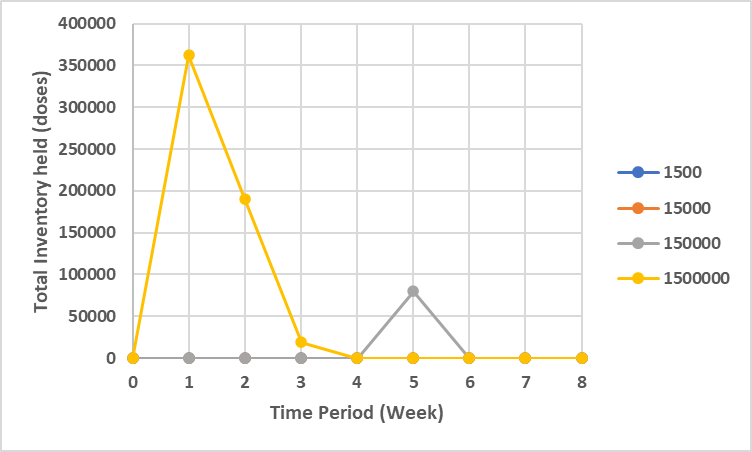}
         \caption{Vaccine inventory held at the SVS.}
         \label{fig:Total inventory held at SVS in doses across the planning horizon}
     \end{subfigure}
     \hfill
     \begin{subfigure}[b]{0.4\textwidth}
         \centering
         \includegraphics[width=\textwidth]{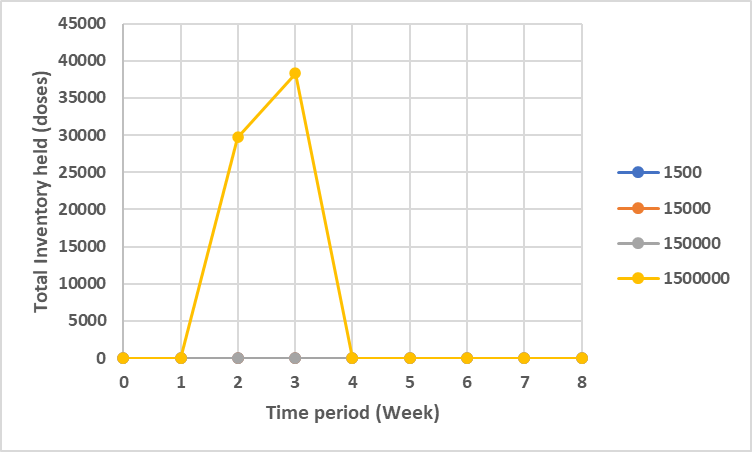}
         \caption{Total vaccine inventory held at the RVSs.}
         \label{fig:Total inventory held at RVSs in doses across the planning horizon}
     \end{subfigure}
     \begin{subfigure}[b]{0.4\textwidth}
         \centering
         \includegraphics[width=\textwidth]{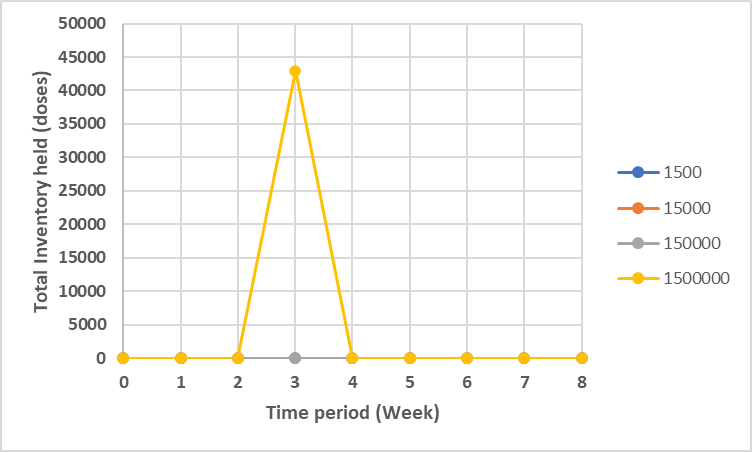}
         \caption{Total vaccine inventory held at the DVSs.}
         \label{fig:Total inventory held at DVS in doses across the planning horizon}
     \end{subfigure}
     \begin{subfigure}[b]{0.4\textwidth}
         \centering
         \includegraphics[width=\textwidth]{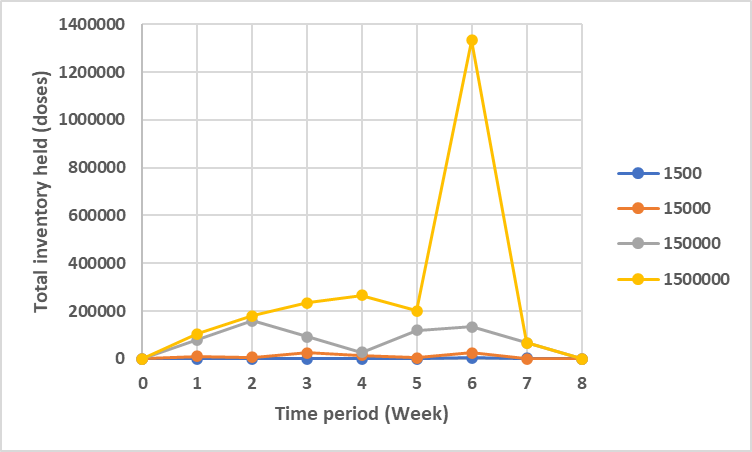}
         \caption{Total vaccine inventory held at the clinics.}
         \label{fig:Total inventory held at Clinics in doses across the planning horizon}
     \end{subfigure}
        \caption{Inventory holding patterns across the planning horizon at each cold chain tier: impact of manufacturer capacity.}
        \label{fig:Variation of inventory levels at different echelons with the manufacturing capacity}
\end{figure}

We conclude this section with a discussion of the manufacturer's capacity on the optimal vaccine distribution patterns across the cold chain. We note before proceeding with the discussion that an 8 week planning horizon was used for these sets of numerical experiments. It has been evident from the COVID-19 vaccination program that the capacity of the manufacturer to meet the demand has been a significant factor in the success of the vaccination program, and hence we examine its impact on logistical considerations across the vaccine cold chain as well. From Figure \ref{fig:Variation of inventory levels at different echelons with the manufacturing capacity}, we observe that when the manufacturing capacity is very low, the optimal solution satisfies the demand of a limited number of clinics, selected because they represent the shortest routes across the cold chain network, to minimize the ordering and the transportation costs. Further, only the demand of the subgroup with the highest priority (elderly recipients) is met due to limited manufacturing capacity. Given the limited manufacturer capacity (especially when the total capacity is less than that required for a subgroup in a single period), instead of transporting vaccines as they become available, vaccine doses are accumulated at the GMSD in each period until the inventory becomes sufficiently high to meet the demand of that subgroup in the limited number of clinics. These are then transported across the cold chain to the clinics without inventory accumulation at the CCPs in the intermediate tiers (i.e., inventory held is zero at all tiers where lead times are zero). Note that the clinics are selected in this situation on the basis of whether they are part of the shortest routes across the cold chain network; however, if clinics are to be prioritized due to some other criteria that become relevant at the time of decision-making (e.g., disease outbreak is high in the catchment area of a particular set of clinics), then any of the associated costs with those clinics (fixed or variable transportation or ordering costs) can be altered to ensure higher priority for the clinics in the catchment area of interest. 

We notice that when the manufacturing capacity is sufficient, the optimal solution yielded by the model holds inventory at lower and intermediate tiers as the ordering and transportation costs from these are significantly lesser than those at higher tiers. Contrary to the case with very low manufacturing capacity, we see from Figures \ref{fig:Total inventory held at Clinics in doses across the planning horizon} and \ref{fig:Total inventory held at RVSs in doses across the planning horizon} that as the manufacturing capacity increases, the inventory levels at the clinics increase from zero while the same decrease to zero at GMSD. The intermediate tiers (SVS, RVS, DVS) hold inventory at initial time periods when the capacity is sufficient to do so, which is depicted in yellow in Figures \ref{fig:Total inventory held at SVS in doses across the planning horizon}, \ref{fig:Total inventory held at RVSs in doses across the planning horizon} and \ref{fig:Total inventory held at DVS in doses across the planning horizon}.

In order to limit the length of the article, we present the computational experiments from the multiple vaccine model in \ref{subsection:multi_vaccine_model}.

%% file: Tables/Single_commodity_COVID.tex
\begin{table}[]
\caption{Ordering, shortage, inventory, and clinic-level vaccination staffing patterns for the lower packed vaccine volume per dose case across the planning horizon.}
\label{table:single-commodity-lower-packed-volume}
\begin{subtable}{\textwidth}
\centering
\caption{Vaccine ordering pattern and associated costs incurred across the planning horizon.} 
\label{table:single-commodity-lower-packed-volume-ordering}
\begin{tabular}{|l|c|c|c|c|c|c|}
\hline
\multicolumn{1}{|c|}{\multirow{2}{*}{\textbf{Route}}}  & \multicolumn{2}{c|}{\textbf{Ordering   Pattern}} & \multicolumn{3}{c|}{\textbf{Transportation Costs (INR)}}  & \multirow{2}{*}{\textbf{Ordering Costs (INR)}} \\ \cline{2-6}
   & \textbf{Week}        & \textbf{Quantity}        & \textbf{Fixed} & \textbf{Variable} & \textbf{Total}   &            \\ \hline
M $\to$ GMSD  & 1 & 333,616 & 40,000 & 14,000 & 54,000 & 200,000 \\ \hline
GMSD $\to$ SVS & 1 & 333,616 & 20,000 & 7,700 & 27,700 & 100,000 \\ \hline
SVS $\to$ RVS 5 & 1 & 333,616 & 12,000 & 1,172 & 13,172 & 75,000 \\ \hline
RVS 5 $\to$ DVS 1 & 1 & 333,616 & 10,000 & 3,465 & 13,465 & 25,000 \\ \hline
DVS 1 $\to$ Clinic 1  & 1 & 26,704 & 5,000 & 0 & 5,000 & 15,000  \\ \hline
DVS 1 $\to$ Clinic 2  & 1 & 26,704 & 5,000 & 0 & 5,000 & 15,000 \\ \hline
DVS 1 $\to$ Clinic 3  & 1 & 26,704 & 5,000 & 638 & 5,638 & 15,000  \\ \hline
DVS 1 $\to$ Clinic 4  & 1 & 26,704 & 5,000 & 424 & 5,424 & 15,000 \\ \hline
DVS 1 $\to$ Clinic 5  & 1 & 26,704 & 5,000 & 0 & 5,000 & 15,000 \\ \hline
DVS 1 $\to$ Clinic 6  & 1 & 26,704 & 5,000 & 413 & 5,413 & 15,000 \\ \hline
DVS 1 $\to$ Clinic 7  & 1 & 26,704 & 5,000 & 649 & 5,649 & 15,000 \\ \hline
DVS 1 $\to$ Clinic 8  & 1 & 26,704 & 5,000 & 363 & 5,363 & 15,000 \\ \hline
DVS 1 $\to$ Clinic 9  & 1 & 26,704 & 5,000 & 0 & 5,000 & 15,000 \\ \hline
DVS 1 $\to$ Clinic 10 & 1 & 26,704 & 5,000 & 499 & 5,499 & 15,000 \\ \hline
DVS 1 $\to$ Clinic 11 & 1 & 11,096 & 5,000 & 5,430 & 10,430 & 15,000 \\ \hline
DVS 1 $\to$ Clinic 12 & 1 & 11,096 & 5,000 & 5,430 & 10,430 & 15,000 \\ \hline
DVS 1 $\to$ Clinic 13 & 1 & 11,096 & 5,000 & 5,705 & 10,705 & 15,000 \\ \hline
DVS 1 $\to$ Clinic 14 & 1 & 11,096 & 5,000 & 587 & 10,587 & 15,000 \\ \hline
DVS 1 $\to$ Clinic 15 & 1 & 11,096 & 5,000 & 5,454 & 10,454 & 15,000 \\ \hline
DVS 1 $\to$ Clinic 16 & 1 & 11,096 & 5,000 & 5,430 & 10,430 & 15,000 \\ \hline
\end{tabular}
\end{subtable}
\end{table}


\begin{table}
\ContinuedFloat
\begin{subtable}{\textwidth}
\centering
\caption{Shortage pattern and associated costs incurred across the planning horizon.}
\label{table:single-commodity-lower-packed-volume-shortage}
\begin{tabular}{|l|c|c|c|c|c|}
\hline
\multirow{2}{*}{\textbf{Clinic}} & \multirow{2}{*}{\textbf{Weeks}} & \multicolumn{3}{c|}{\textbf{Number of shortages}} & \multirow{2}{*}{\textbf{Total Shortage   Cost (INR)}} \\ \cline{3-5}
&    & \textbf{Children}     & \textbf{Adults}      & \textbf{Elderly}     &   \\ \hline
1-10 & 1,2 & 3,004 & 3,204  & 468 & 515,671,968      \\ \hline
11-16  & 1,2 & 1,248 & 1,331 & 195 & 214,264,288        \\ \hline
\end{tabular}
\end{subtable}
\end{table}


\begin{table}
\ContinuedFloat
\begin{subtable}{\textwidth}
\centering
\caption{Inventory holding patterns and the associated costs incurred across the planning horizon.} 
\label{table:single-commodity-lower-packed-volume-inventory}
\begin{tabular}{|l|c|c|c|}
\hline
\multicolumn{1}{|c|}{\textbf{CCP} } & \textbf{Weeks} & \textbf{Number of Inventory units held}  & \textbf{Total Inventory cost (INR)} \\ \hline
GMSD, SVS, RVS, DVS & All & 0 & 0  \\ \hline
Clinics 1-10 & 2; 3; 4; 5 & 26,704; 20,028; 13,352; 6,676 & 20,028 \\ \hline
Clinics 11-16 & 2; 3; 4; 5 & 11,096; 8,322; 5,548; 2,774 & 8,322 \\ \hline
\end{tabular}
\end{subtable}
\end{table}


\begin{table}
\ContinuedFloat
\begin{subtable}{\textwidth}
\centering
\caption{Vaccination staffing pattern across the planning horizon: number of health workers hired, fired, and total staff numbers at the clinics.}
\label{table:single-commodity-lower-packed-volume-staff-planning}
\begin{tabular}{|l|c|c|c|c|}
\hline
\textbf{Clinic} & \textbf{Weeks} & \textbf{Number hired} & \textbf{Number fired} & \textbf{Staff numbers} \\ \hline
1-10 & 3,4,5,6 & 10,0,0,0 & - & 10,10,10,10    \\ \hline
11-16 & 3,4,5,6 & 5,0,0,0  & - & 5,5,5,5        \\ \hline
\end{tabular}
\end{subtable}
\end{table}

%% file: Tables/Single_commodity_measles.tex
\begin{table}[]
\caption{Ordering, shortage, inventory, and clinic-level vaccination staffing patterns for the higher packed vaccine volume per dose case across the planning horizon.}
\label{table:single-commodity-higher-packed-volume}
\begin{subtable}{\textwidth}
\centering
\caption{Ordering pattern and associated costs incurred across the planning horizon.} 
\label{table:single-commodity-higher-packed-volume-ordering}
\begin{tabular}{|l|c|c|c|c|c|c|}
\hline
\multicolumn{1}{|c|}{\multirow{2}{*}{\textbf{Route}}}  & \multicolumn{2}{c|}{\textbf{Ordering   Pattern}} & \multicolumn{3}{c|}{\textbf{Transportation Costs (INR)}}  & \multirow{2}{*}{\textbf{Ordering Costs (INR)}} \\ \cline{2-6}
   & \textbf{Week}        & \textbf{Quantity}        & \textbf{Fixed} & \textbf{Variable} & \textbf{Total}   &            \\ \hline
M $\to$ GMSD  & 1 & 333,616 & 40,000 & 14,000 & 54,000 & 200,000 \\ \hline
GMSD $\to$ SVS & 1 & 333,616 & 20,000 & 7,700 & 27,700 & 100,000 \\ \hline
SVS $\to$ RVS 5 & 1 & 171,320 & 12,000 & 1,172 & 13,172 & 75,000 \\ \hline
SVS $\to$ RVS 6 & 1  & 162,296 & 12,000 & 1,248 & 13,248 & 75,000 \\ \hline
RVS 5 $\to$ DVS 1 & 1 & 171,320 & 10,000 & 3,465 & 13,465 & 25,000 \\ \hline
RVS 6 $\to$ DVS 2 & 1 & 162,296 & 10,000 & 1,300 & 11,300 & 25,000 \\ \hline
DVS 1 $\to$ Clinic 1  & 1 & 26,704 & 5,000 & 0 & 5,000 & 15,000  \\ \hline
DVS 1 $\to$ Clinic 2  & 1 & 26,704 & 5,000 & 0 & 5,000 & 15,000 \\ \hline
DVS 1 $\to$ Clinic 5  & 1 & 26,704 & 5,000 & 0 & 5,000 & 15,000 \\ \hline
DVS 1 $\to$ Clinic 8  & 1 & 26,704 & 5,000 & 363 & 5,363 & 15,000 \\ \hline
DVS 1 $\to$ Clinic 9  & 1 & 26,704 & 5,000 & 0 & 5,000 & 15,000 \\ \hline
DVS 1 $\to$ Clinic 10 & 1 & 26,704 & 5,000 & 499 & 5,499 & 15,000 \\ \hline
DVS 1 $\to$ Clinic 15 & 1 & 11,096 & 5,000 & 5,454 & 10,454 & 15,000 \\ \hline
DVS 2 $\to$ Clinic 3  & 1 & 26,704 & 5,000 & 5,344 & 10,344 & 15,000 \\ \hline
DVS 2 $\to$ Clinic 4  & 1 & 26,704 & 5,000 & 5,030 & 10,030 & 15,000 \\ \hline
DVS 2 $\to$ Clinic 6  & 1 & 26,704 & 5,000 & 5,248 & 10,248 & 15,000 \\ \hline
DVS 2 $\to$ Clinic 7  & 1 & 26,704 & 5,000 & 4,768 & 9,768 & 15,000 \\ \hline
DVS 2 $\to$ Clinic 11 & 1 & 11,096 & 5,000 & 0 & 5,000 & 15,000 \\ \hline
DVS 2 $\to$ Clinic 12 & 1 & 11,096 & 5,000 & 0 & 5,000 & 15,000 \\ \hline
DVS 2 $\to$ Clinic 13 & 1 & 11,096 & 5,000 & 275 & 5,275 & 15,000 \\ \hline
DVS 2 $\to$ Clinic 14 & 1 & 11,096 & 5,000 & 232 & 5,232 & 15,000 \\ \hline
DVS 2 $\to$ Clinic 16 & 1 & 11,096 & 5,000 & 0 & 5,000 & 15,000 \\ \hline  
\end{tabular}
\end{subtable}
\end{table}


\begin{table}
\ContinuedFloat
\begin{subtable}{\textwidth}
\centering
\caption{Shortage pattern and associated costs incurred across the planning horizon.}
\label{table:single-commodity-higher-packed-volume-shortage}
\begin{tabular}{|l|c|c|c|c|c|}
\hline
\multirow{2}{*}{\textbf{Clinic}} & \multirow{2}{*}{\textbf{Weeks}} & \multicolumn{3}{c|}{\textbf{Number of shortages}} & \multirow{2}{*}{\textbf{Total Shortage   Cost (INR)}} \\ \cline{3-5}
&    & \textbf{Children}     & \textbf{Adults}      & \textbf{Elderly}     &   \\ \hline
1-10 & 1,2 & 3,004 & 3,204  & 468 & 515,671,968      \\ \hline
11-16  & 1,2 & 1,248 & 1,331 & 195 & 214,264,288        \\ \hline
\end{tabular}
\end{subtable}
\end{table}


\begin{table}
\ContinuedFloat
\begin{subtable}{\textwidth}
\centering
\caption{Inventory holding pattern and associated costs incurred costs incurred across the planning horizon.} 
\label{table:single-commodity-higher-packed-volume-inventory}
\begin{tabular}{|l|c|c|c|}
\hline
\multicolumn{1}{|c|}{\textbf{CCP}} & \textbf{Weeks} & \textbf{Number of Inventory units held}  & \textbf{Total Inventory cost (INR)} \\ \hline
GMSD, SVS, RVS, DVS & All & 0 & 0  \\ \hline
Clinics 1-10 & 2; 3; 4; 5 & 26,704; 20,028; 13,352; 6,676 & 20,028 \\ \hline
Clinics 11-16 & 2; 3; 4; 5 & 11,096; 8,322; 5,548; 2,774 & 8,322 \\ \hline
\end{tabular}
\end{subtable}
\end{table}


\begin{table}
\ContinuedFloat
\begin{subtable}{\textwidth}
\centering
\caption{Vaccination staffing pattern across the planning horizon: number of health workers hired, fired, and total staff numbers at the clinics.}
\label{table:single-commodity-higher-packed-volume-staff-planning}
\begin{tabular}{|l|c|c|c|c|}
\hline
\textbf{Clinic} & \textbf{Weeks} & \textbf{Number hired} & \textbf{Number fired} & \textbf{Staff numbers} \\ \hline
1-10 & 3,4,5,6 & 10,0,0,0 & - & 10,10,10,10    \\ \hline
11-16 & 3,4,5,6 & 5,0,0,0  & - & 5,5,5,5        \\ \hline
\end{tabular}
\end{subtable}
\end{table}

%% file: Sections/Pre_processing_results.tex
The formulations that we propose in this paper support making a multitude of decisions optimally, ranging from optimal routing and scheduling decisions for the cold chain network to vaccine recipient group selection and vaccination staffing decisions. The results discussed in Sections \ref{subsection:single_vaccine_model} and \ref{subsection:multi_vaccine_model} are generated for relatively small instances of the cold chain network, and are meant primarily to illustrate (a) how the output of various models within our framework can be organized, and (b) the types of analyses that can be performed using our decision support framework. However, realistic instances of the cold chain network are likely to be significantly larger. Thus, it is important to analyze the computational expense of the optimization formulations within our framework for more realistic instances of the cold chain network given the multitude of decisions that we attempt to inform through our models. 

We begin by considering the computational cost of solving the single vaccine model for the two cases presented in Section \ref{subsection:single_vaccine_model} for the entire state of Bihar: the base case (with the lower packed volume per dose) and the case with the higher packed volume per dose vaccine. This yields a problem with 1 SVS, 9 RVSs, 38 DVSs and a total of 606 clinics. We use the Gurobi commercial solver with its Python programming application process interface to solve the problem. All computational analyses were run on a workstation with an Intel R Core \textit{i}5-7200 processor with a clock speed of 2.50 gigaHertz and 8 gigabytes of memory. As part of these analyses, we define the following measure, referred to as the MIP gap, of how close the solution generated by the solver is to the dual objective bound for the optimization problem (i.e., a lower bound for minimization integer programs).
\begin{equation*}
    \text{MIP gap} = \frac{\lvert Z_{P}-Z_{D} \rvert}{\lvert Z_{P} \rvert}
\end{equation*}

Here $Z_P$ is the primal objective bound (i.e., the incumbent objective function value), and $Z_D$ is the dual objective bound obtained after 42000s. The MIP gap can be used to specify a termination criterion for the algorithm; for example, if set to 1\%, it implies that the algorithm will terminate if the primal objective function value is within 1\% of the dual objective bound. Thus in the subsequent discussion, we compare the computational expense of the formulation in terms of the computational runtime required to achieve an MIP gap less than some prespecified threshold (e.g., 1\%). 

Solving the single vaccine model for the entire state of Bihar for the base case required 95 seconds to yield an MIP gap of 0.0032\%; however, when the larger packed volume per dose was used, the solver did not reach an MIP gap below 1.12\% even after 10,000 seconds. This is not unexpected because, as discussed in Section \ref{subsection:single_vaccine_model}, when the transportation and storage capacities are sufficiently high (the base case analysis), the solver just selects the shortest route to the clinics each time it ships vaccines during the planning horizon. However, when these capacities are lower (as in the higher packed volume per dose case), the algorithm has to consider ordering from DVSs other than those in the district where a given clinic is located in order to satisfy the demand associated with a given clinic. It is in these situations that employing preprocessing techniques to reduce computational runtimes becomes relevant. 

We remind readers here that the vaccine cold chain in India consists of the following tiers, in order of flow: vaccine manufacturers, GMSDs (national level tier), SVSs (state level tier), RVSs (regional level tier) to DVSs (district level tier), clinics or vaccination centers (where vaccine administration actually occurs). In order to gain a preliminary understanding of the basis on which routes are selected through this network, we consider a simpler problem by removing the RVS tier entirely and analyse the optimal solution generated for such a network. Note that the removal of a tier in this manner is not an unrealistic operation, and has support in the literature \citep{assi2013removing}; in situations such as the COVID-19 pandemic where the vaccination rate is crucial, public health authorities may elect to speed up the vaccine ordering and delivery process by skipping intermediate tiers. We anticipate developing a decision support software tool utilizing this framework which allows removal or inclusion of as many tiers as required for the actual vaccine distribution process. The optimization problem formulation will also be automatically selected based on the tiers included in the model. 

As discussed above, clinics do not necessarily order from the DVS in the district where they are located when transportation and storage capacities are limited. `Cross-ordering' can happen between districts due to the interplay of storage and transportation capacities, ordering and transportation costs. For example, a DVS in a district with lower demand may, in addition to satisfying the demand of its own district, also satisfy the demand of another district with a higher demand. This often occurs because of the limit on the capacity of the vehicles used to transport vaccines from the SVS to the DVSs, which implies that the DVS in the district with higher demand may need an additional order from the SVS, perhaps even in a previous time period, to fully satisfy the demand of the clinics in its district in a given time period. This would imply that additional fixed ordering and fixed transportation costs are incurred (associated with vaccine delivery from the SVS to the DVS), and in situations where these are higher than the sum of the corresponding costs and also the transportation costs from a DVS in another district to the clinic under consideration, cross-ordering occurs. 

\begin{figure}[h]
    \centering
    \includegraphics[width=5in]{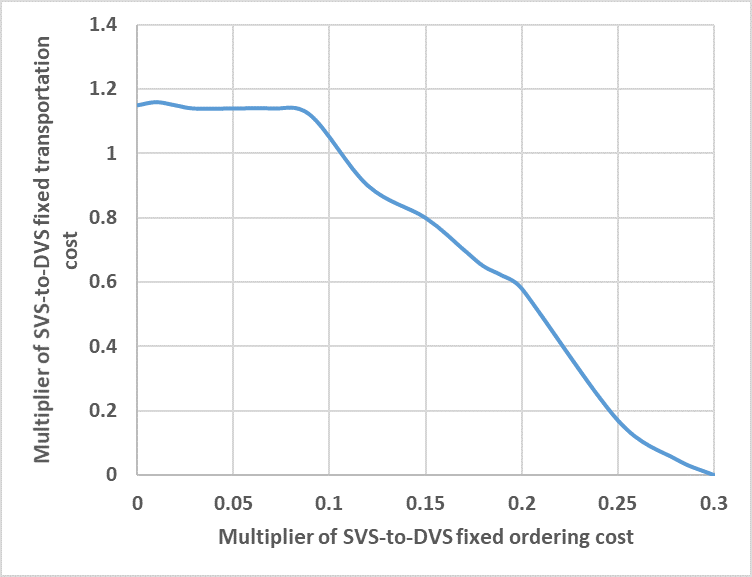}
    \caption{Threshold points below which cross ordering stops}
    \label{fig:Threshold points below which cross ordering stops}
\end{figure}

The benefit of preventing cross-ordering from the perspective of computational expense is clear from the above analysis. Thus, we constructed a relationship between the fixed transportation cost and the ordering cost from the SVS to the DVS by decreasing these costs from their current estimates and determining whether cross-ordering occurs at each combination of estimates of these costs. In Figure~\ref{fig:Threshold points below which cross ordering stops}, we depict this relationship. On the Y-axis, we have the multipliers of the fixed transportation cost from the SVS to the DVS (for example, a value of 0.5 implies the fixed transportation cost is half the current estimate), and on the X-axis, we have the multipliers of the corresponding fixed ordering cost. A point on the graph implies that at combinations of these fixed costs below this point, cross-ordering does not occur. For example, the point (0.15, 0.8) on the graph indicates that at fixed ordering and transportation costs that are simultaneously less than 15\% and 80\% of the current estimates, respectively, cross-ordering does not occur. As discussed earlier, if these fixed costs are taken to be proxies of the efficiency of the vaccine ordering process, then the above result implies increasing the efficiencies of the ordering process to points below the graph would yield intuitive vaccine distribution patterns across the cold chain network.

The above analysis is used to illustrate how our computational framework can be used to determine relationships such as that depicted in Figure~\ref{fig:Threshold points below which cross ordering stops} for the specific instance of the cold chain of interest to the analyst. Deriving such a relationship would help reduce computational expense of the framework, and also yield more intuitive vaccine distribution patterns that may in turn help motivate efforts to make ordering processes more efficient. 


We now consider other types of preprocessing that do not require changes to model parameter estimates. To illustrate the use of one of these techniques, we continue with the higher packed volume per dose case for the subsequent analyses, but without lead times at any of the tiers. We begin by observing the ordering pattern in the solution generated by running the solver for 42,000 seconds for this instance of the problem. We see that all 606 clinics order from the four nearest districts to them. This suggests that adding constraints to this effect - that clinics order from the four nearest districts to them - could reduce the number of cuts in the branch and cut algorithm employed by the solver and yield significantly faster computational runtimes such that the MIP gap threshold is breached. 

\begin{figure}
    \centering
    \includegraphics[width=5.3in,height=3.2in]{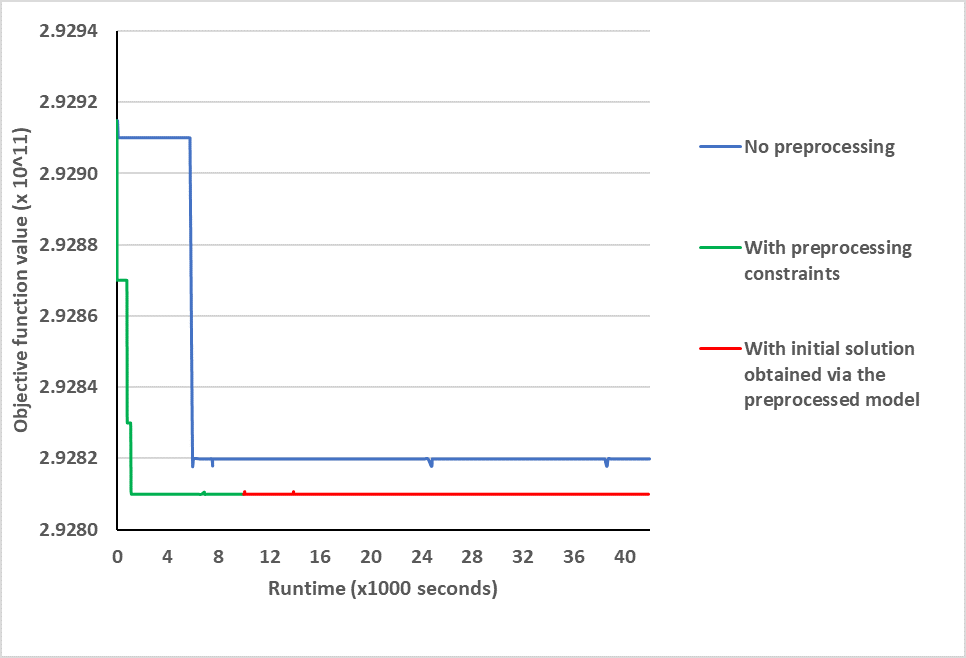}
    \caption{Objective function trajectories by time with and without preprocessing.}
    \label{fig:Variation of objective function values with program running time}
\end{figure}

Therefore, we add preprocessing constraints that restrict the clinics to order only from their four nearest DVSs. Then, we specify the solution obtained by running this preprocessed model for 10,000 seconds as the initial solution to the original instance of the problem without preprocessing constraints and run the solver for 32,000 seconds. The trajectories of the objective function values against the runtime observed in each of the following three cases is depicted in Figure \ref{fig:Variation of objective function values with program running time}: 
\begin{enumerate}
    \item No preprocessing
    \item With preprocessing constraints 
    \item With initial solution obtained via the preprocessed model
\end{enumerate}

We see from Figure~\ref{fig:Variation of objective function values with program running time} that the objective function value for the preprocessed version of the model is lower than that from the version without preprocessing, and that it is reached significantly faster (in around 1,000 seconds) than the version without preprocessing. The same observation holds for the version with the initial solution obtained from the preprocessed version also. The MIP gap for the model with the preprocessing constraints is also significantly lower than those for the other models: s0.012\% versus 0.0197\% and 0.0181\% for the model with no preprocessing and the model with the initial solution from preprocessing, respectively.


We now use these insights for the realistic problem instance discussed in the beginning of this section - the higher packed volume per dose case for the entire state of Bihar, with lead times - which had not breached the MIP gap threshold of 1\% even after 10,000 seconds. First, we apply preprocessing constraints that restrict each clinic from ordering from its nearest $n$ districts. We then varied $n$ to determine how it impacts the time taken to breach the MIP gap threshold. We find that when $n = 6$, the MIP gap reaches 0.69\% in 4,279 seconds. When $n = 8$ or $n = 10$, the MIP gap is not breached even after 10,000 seconds; however, when $n = 12 $, the MIP gap reaches 0.47\% in 1,008 seconds.

Finally, we specify the solution obtained from adding preprocessing constraints restricting each clinic to order from its nearest 12 DVSs as the initial solution to the problem without preprocessing constraints. We find that the MIP gap reaches 0.04\% in 1,481 seconds.

The above analyses indicate how the computational expense associated with the models in our framework can be reduced by applying one or more of the preprocessing techniques discussed above.

\subsection{Robust Version of the Single Vaccine Formulation: Computational Experiments}
\label{subsection:robustresults}
In this section, we present the analyses pertaining to the robust optimization version of the single vaccine model. We modeled uncertainty in ordering costs, holding costs, demand, manufacturing capacity and cold chain capacity at the clinic level. Ordering costs and inventory holding costs are typically not straightforward to estimate \citep{hopp2011factory}. Hence we analyze the impact of uncertainty in these parameters on the optimal solution of the single vaccine model. In the following analyses, we compare the level of conservativeness of the deterministic formulation and robust formulation with box and budgeted uncertainty sets using randomly generated data instances.

More specifically, we compare the performance of the deterministic formulation with the robust formulations with box and budgeted uncertainty sets with data instances generated randomly based on the deviations considered in the robust formulations. We generate 10 instances with low levels of uncertainty, 10 with medium levels of uncertainty and 10 with high levels of uncertainty and substitute it in the optimal network created by the deterministic and robust formulations. In the case of low uncertainty, the parameter data is allowed to deviate between 0 to 50\% of the original deviations considered in the uncertainty sets. For the case of medium uncertainty and high uncertainty data instances the corresponding figures are 50\% to 100\% and 100\% to 150\% respectively. The results of the experiments are presented in Tables~\ref{table12a}, \ref{table12b} and \ref{table12c} in \ref{aprobust}. In the table, the objective function values of the deterministic case and the robust formulations with box uncertainty and with budgeted uncertainty sets, respectively, are presented in columns 2, 3 and 4. The entries in the table show that though the performances of the robust formulations on estimated deterministic parameters are poor when compared to the deterministic case (which is optimized for this data), the robust formulations provide cost savings in all the randomly generated data instances irrespective of the level of uncertainty. The magnitude of cost savings achieved through the formulation with box uncertainty set over the deterministic formulation varies between INR 0.36 million to INR 0.64 million in the case of low uncertainty, INR 0.27 million to INR 0.81 million in case of medium uncertainty and INR 0.24 million to 0.87 mn INR in case of high uncertainty. This is given in the column 4 of the Tables \ref{table12a}, \ref{table12b} and \ref{table12c} respectively. In column 5 of the tables we provide the cost savings obtained through the robust formulation with budgeted uncertainty set over the robust formulation with box uncertainty set over varying levels of uncertainty. The results show that with a carefully selected budget, the decision maker can save significant cost while using robust formulation with budgeted uncertainty sets. The cost saving varies between INR 0.56 million to INR 0.91 million for low uncertainty, INR 0.39 million to INR 1.04 million for medium uncertainty and INR 0.40 million to INR 1.20 million for high uncertainty cases. 

Thus the above results demonstrate the advantage of the robust formulations when the model parameter estimates are subject to uncertainty. The experiments also suggest the advantage of incorporating uncertainty modeling for the decision makers under limited budgets available for the vaccination exercise.

\section{Discussion \& Conclusions}
\label{disc}
We present in this paper a decision support framework for optimal vaccine distribution across the vaccine cold chain network. We present two integer linear programming models within the framework: a model considering a single vaccine for a single disease, and a model considering multiple vaccines for a single disease. The model can be extended to incorporate multiple vaccines for multiple diseases as well. Finally, we develop two robust optimization formulations for the single vaccine model: one with box uncertainty sets and the other with budgeted uncertainty sets. As discussed in Section~\ref{section:literature}, a key advantage of our optimal vaccine distribution framework is that it only contains linear integer programming models in comparison with previous work, facilitating easier generation of exact solutions. 

The single and multiple vaccine models support a multitude of decisions, ranging from routing and scheduling of vaccine deliveries across the cold chain network within the planning horizon to hiring and firing decisions for vaccination staff at each clinic. We also consider every single tier within the vaccine cold chain network described by the Government of India \citep{NCCVMRC}. However, we understand that not all these decisions may be relevant for a public health planning authority when planning vaccine distribution across the cold chain. Therefore thus, we intend to incorporate this framework within a computational decision support tool in a modular form. This implies that the tool will allow an analyst to choose, for instance, whether to use the single vaccine or the multiple vaccine model; or to use the robust versions of these formulations. Further, the analyst will have the option of deciding whether to include or exclude a particular cold chain tier from their analysis. For example, as discussed in Section \ref{subsection:pre_processing}, the analyst can choose to exclude the regional vaccine store from their analysis, and the appropriate formulation will then be selected in the back-end of the decision support tool. The analyst will also be able to choose whether to include or exclude certain sets of decisions in their analysis: for example, they may choose to exclude vaccination staffing decisions, in which case the appropriate terms will automatically be omitted from the formulation in the back-end. Preprocessing can also be included automatically to a certain extent; for example, if generating the solution to the analyst's instance of the model requires more than some prespecified threshold runtime, then preprocessing techniques such as those discussed in Section \ref{subsection:pre_processing} can automatically be deployed. 

A key limitation of the framework that we propose involves the multitude of parameters that need to be estimated for each model within the framework. For example, an analyst may not have access to primary data required to estimate inventory holding costs for a vaccine at every cold chain point within the network. However, we note that such data may be available to public health authorities overseeing vaccine distribution; hence for analysts working in these organizations, estimating these parameters may be a one-time effort. Further, as discussed in the preceding sections, it is the relative value of these parameters that are more relevant than actual estimates themselves. For example, if it is known that the ordering process at a particular cold chain tier, or a particular group of facilities within a cold chain tier (e.g., clinics in a particular district), is significantly more efficient than at comparable tiers or facilities, then the ordering costs for these facilities can be adjusted (starting from a reasonable initial estimate) to reflect the disparity in efficiency in their respective ordering processes. 

Overall, given the ubiquity of multi-tier vaccine cold chain networks within public health systems across the world (as discussed in Section 1), we believe that the decision support framework that we propose in this study can be useful for stakeholders within public health planning authorities in these health systems.

%% file: Sections/Mathematical_Formulation_Multi_Commodity.tex
In our proposed multiple vaccine formulation (an extension of the single vaccine formulation in Section~\ref{subsection:single_vaccine_formulation}), the cold chain points are indicated by the same indices as for the single vaccine formulation. However, we have an additional index for the vaccine type. 
\begin{itemize}
  \item Vaccine index, $k \in \{1,2,3......K\} \; (K \leq M)$
\end{itemize}

\input{Formulation/parameters_multi}
\begin{flushleft}
    \normalsize \textbf{Total demand}
\end{flushleft}
The following is the total demand by all the subgroups at all the clinics across the entire planning horizon. This is the same as that for single vaccine formulation.

$Q$ = $\mathop{\sum_{t}^{T}\sum_{i}^{I}\sum_{j}^{J}}$ $\frac{D^{ij}_{t}}{w}$

\input{Formulation/decision_variables_multi}
\input{Formulation/objective_multi}
\input{Formulation/constraints_multi}

%% file: Formulation/parameters_multi.tex
\begin{flushleft}
    \normalsize \textbf{Parameters}
\end{flushleft}
\begin{itemize}
    \item Inventory holding cost parameters for vaccine $k$ at each cold chain point: $h_t^{kg}$, $h_t^{ks}$, $h_t^{kr}$, $h_t^{kd}$, $h_t^{ki}$. Units: INR/dose/week.
    
    
    
    
    
  \item Transportation cost (fixed + variable) per vehicle transporting vaccines from one CCP to the next lower-tier CCP (e.g., manufacturer $m$ to GMSD $g$, or DVS $d$ to clinic $i$): $K^{mg}$, $K^{gs}$, $K^{sr}$, $K^{rd}$, $K^{di}$. Units: INR/vehicle.
    \item Vaccine inventory holding capacity at each CCP: $B_t^g$, $B_t^s$, $B_t^r$, $B_t^d$, $B_t^i$. Units: $cm^3$.
\item Capacity of each vehicle transporting  vaccines from one CCP to the next lower-tier CCP (e.g., manufacturer $m$ to GMSD $g$, or DVS $d$ to clinic $i$): $C_t^{mg}$, $C_t^{gs}$, $C_t^{sr}$, $C_t^{rd}$, $C_t^{di}$. Units: $cm^3$.
\item Fixed cost (ordering) of ordering vaccine by a CCP from the next higher-tier CCP (e.g., GMSD $g$ from manufacturer $m$, or clinic $i$ from DVS $d$): $S_t^{kmg}$, $S_t^{gs}$, $S_t^{sr}$, $S_t^{rd}$, \textbf{$S_t^{di}$}. Units: INR/delivery. Note that the parameters $S^{kmg}_t$ alone are indexed by vaccine type (the index $k$) along with manufacturer, GMSD and time indices. This is done to account for the possibility that a manufacturer may supply more than one vaccine and the fact that different vaccines may have different ordering processes at the interface between the manufacturer and the government, and hence may incur different ordering costs.
\item Maximum number of vehicles available for transportation from one CCP to the next lower-tier CCP (e.g., manufacturer `m' to GMSD $g$, or DVS $d$ to clinic $i$): $N^{mg}$, $N^{gs}$, $N^{sr}$, $N^{rd}$, $N^{di}$. Units: vehicles.
\item Lead times of delivery of vaccines from one CCP to the next lower-tier CCP (e.g., manufacturer $m$ to GMSD $g$, or DVS $d$ to clinic $i$): $L^{mg}$, $L^{gs}$, $L^{sr}$, $L^{rd}$, $L^{di}$. Units: weeks.
    \item Lead times of administration of vaccines at the clinics: $L^{ij}$. Units: weeks. 
  
    
    
    
    
    
    
    \item Initial inventory of vaccine $k$ held at the CCPs: $I^{kg}_{0}$, $I^{ks}_{0}$, $I^{kr}_{0}$, $I^{kd}_{0}$, $I^{ki}_{0}$ Units: doses.
  
    
    
    
    
    

  \item Miscellaneous parameters:
  \begin{itemize}
  
    \item Demand by sub-group $j$ in the clinic $i$ at time $t$ (in doses/week): \hfill $D^{ij}_{t}$
    
    \item Shortage cost (INR) of not vaccinating a customer in subgroup $j$ at time $t$\hfill $P^j_{t}$
    
    \item Clinical services cost per customer in subgroup $j$ (e.g., INR/dose) for vaccine $k$: \hfill $V^{kj}$
    
    \item Average time (e.g., minutes/dose) required for administration of one vaccine dose: \hfill $T_o$
    
    \item Availability of a health worker in hours at clinic $i$ for time period $t$ (e.g., 40 hours/week): \hfill $N_t^i$
    
    
    \item The production capacity of manufacturer $m$ for vaccine $k$ at time $t$ (in doses): \hfill $B_t^{mk}$
    
    \item Wastage factor (proportion of each dose wasted; that is, effective volume required per dose is $\frac{1}{w}$): \hfill $w$
    
    \item Wages of health workers per time period (e.g., INR/week): \hfill $L$
    
    \item Fixed cost (INR) of hiring one health worker: \hfill $E$
    
    \item Fixed cost (INR) of firing one health worker: \hfill $F$
    
    \item Packed vaccine volume per dose ($cm^3$/dose) of vaccine $k$: \hfill $P_{k}$
    
    \item Probability of getting exposed to SARS-CoV2 \hfill $p$
    
    \item Probability of developing the disease after vaccination by vaccine $k$ upon exposure to SARS-CoV2 \hfill $\eta_k$
    
  \end{itemize}

\end{itemize}

%% file: Formulation/decision_variables_multi.tex
\begin{flushleft}
    \normalsize \textbf{Decision Variables}
\end{flushleft}
\begin{itemize}
    \item Number of doses of vaccine $k$ held at each CCP at the end of time $t$: $I^{kg}_t$, $I^{ks}_t$, $I^{kr}_t$, $I^{kd}_t$, $I^{ki}_t$
      
      
        
        
        
        
    \item Number of doses of vaccine $k$ delivered from one CCP to the next lower-tier CCP at the beginning of time $t$ (e.g., from manufacturer $m$ to GMSD $g$, or from DVS $d$ to clinic $i$): $q^{kmg}_t$, $q^{kgs}_t$, $q^{ksr}_t$, $q^{krd}_t$, $q^{kdi}_t$
      
        
        
        
        
        
    \item Number of vehicles required for transporting vaccines from one CCP to the next lower-tier CCP  at time $t$ (e.g., manufacturer $m$ to GMSD $g$, or DVS $d$ to clinic $i$): $n_t^{mg}$, $n_t^{gs}$, $n_t^{sr}$, $n_t^{rd}$, $n_t^{di}$.
      
        

        
        
        
    \item Binary assignment variable indicating whether an order has been placed from a CCP by the next lower-tier CCP (e.g., order placed by GMSD $g$ from manufacturer $m$): $x^{kmg}_t$, $x^{gs}_t$, $x^{sr}_t$, $x^{rd}_t$, $x^{di}_t$. Note that, similar to the vaccine ordering costs, the decision variables $x^{kmg}_t$ alone are indexed by vaccine type (the index $k$) along with manufacturer, GMSD and time indices.
      
        
        
        
        
        
    \item Number of persons not vaccinated (i.e., number of shortages) and number of doses of vaccine $k$ administered in subgroup $j$ in clinic $i$ at time $t$, respectively: $s^{ij}_{t}$, $w^{ijk}_{t}$
      
      
      
    \item Number of health workers working, hired and fired in clinic $i$ at time $t$ respectively: $n^i_t$, $h^i_t$, $f^i_t$
      
      
      

  \end{itemize}

%% file: Formulation/objective_multi.tex
\begin{flushleft}
    \normalsize \textbf{Objective Function}
\end{flushleft}
\textbf{Min $J$} = \\*

$\mathop{\sum_{t}\sum_{m}\sum_{g}} K^{mg} n_t^{mg} +
\mathop{\sum_{t}\sum_{g}\sum_{s}} K^{gs} n_t^{gs} +
\mathop{\sum_{t}\sum_{s}\sum_{r}} K^{sr} n_t^{sr} +
\mathop{\sum_{t}\sum_{r}\sum_{d}} K^{rd} n_t^{rd} +
\mathop{\sum_{t}\sum_{d}\sum_{i}} K^{di} n_t^{di} +\\*$

\fbox{Transportation cost $\times$ no of trucks from one cold chain point to other}\\*

$\mathop{\sum_{t}\sum_{k}\sum_{g}} h_t^{kg} I_t^{kg}+
\mathop{\sum_{t}\sum_{k}\sum_{s}} h_t^{ks} I_t^{ks}+
\mathop{\sum_{t}\sum_{k}\sum_{r}} h_t^{kr} I_t^{kr}+
\mathop{\sum_{t}\sum_{k}\sum_{d}} h_t^{kd} I_t^{kd}+ \mathop{\sum_{t}\sum_{k}\sum_{i}} h_t^{ki} I_t^{ki}+\\*$

\fbox{Holding cost $\times$ Inventory at cold chain points}\\*

$\mathop{\sum_{t}\sum_{i}\sum_{j}} p~P^j_{t} s^{ij}_{t}$ + $\mathop{\sum_{t}\sum_{k}\sum_{i}\sum_{j}} \left(1-\eta_{k}\right) p~P^j_{t}~ w^{ijk}_{t}\\*$

\fbox{Shortage cost associated with getting disease without and with  vaccination }\\*

$\mathop{\sum_{t}\sum_{k}\sum_{i}\sum_{j}} V^k_j w^{ijk}_{t} +\\*$

\fbox{Clinical cost $\times$ Consumption of vaccine units at clinic }\\*

$\mathop{\sum_{t}\sum_{k}\sum_{m}\sum_{g}} S_t^{kmg} x_t^{kmg}+
\mathop{\sum_{t}\sum_{g}\sum_{s}} S_t^{gs} x_t^{gs}+
\mathop{\sum_{t}\sum_{s}\sum_{r}} S_t^{sr} x_t^{sr}+
\mathop{\sum_{t}\sum_{r}\sum_{d}} S_t^{rd} x_t^{rd}+
\mathop{\sum_{t}\sum_{d}\sum_{i}} S_t^{di} x_t^{di} +\\*$

\fbox{Ordering cost (cost incurred if even one vaccine unit is ordered)}\\*

$\mathop{\sum_{t}\sum_{i}} L n^i_t +
\mathop{\sum_{t}\sum_{i}} E h^i_t +
\mathop{\sum_{t}\sum_{i}} F f^i_t$\\

\fbox{Labour costs (Wages, hiring and firing costs)}

%% file: Formulation/constraints_multi.tex
\begin{flushleft}
    \normalsize \textbf{Subject to constraints:}
\end{flushleft}
\begin{itemize}
        \setcounter{equation}{0}
        
        \item Production capacity constraints associated with the manufacturers.
        \begin{equation}
     \mathop{\sum_{g=1}^{G} q^{kmg}_t \leq B^{mk}_t} ~~\forall~ k,m,t
        \end{equation}
    
    \item Facility selection constraints.
        \begin{equation}
            \mathop{\sum_{a}} x_t^{ab} \leq 1  ~~ \forall~ b,t ~~~~\text{if } a \in \{g,s,r,d\} ~\text{then}~ b \text{ := the next lower-tier value from } \{s,r,d,i\}
        \end{equation}
        For example,
        \begin{equation*}
            \mathop{\sum_{g}} x_t^{gs} \leq 1  ~~ \forall~ s,t 
        \end{equation*}
        
    
    \item To ensure consistency of $x$ and $q$.
        \begin{equation}
                \sum_{k=1}^{K} q_t^{kab} P_{k} \leq ( N^{ab} C^{ab}_t )~ x_t^{ab} ~~\forall~ a,b,t 
            \end{equation}
         Here, if a $\in$ \{m,g,s,r,d\}, then b := the next lower-tier value from \{g,s,r,d,i\}. For example,
          \begin{equation*}
            q_t^{kmg} \leq ( N^{mg} C^{mg}_t )~ x_t^{kmg} ~~\forall~ k, m,g,t
            \end{equation*}
    
        
        \begin{equation*}
            \sum_{k=1}^{K} q_t^{kgs} P_k \leq ( N^{gs} C^{gs}_t )~ x_t^{gs} ~~\forall~ g,s,t
        \end{equation*}
    
       
       
    
    \item Number of vehicles. 
        \begin{equation}
            \frac{\sum_{k=1}^{K} q_t^{kab} P_k}{C_t^{ab}} \leq  n_t^{ab} ~~\forall~ a,b,t 
            \end{equation}
            Here, if a $\in$ \{m,g,s,r,d\} then b := the next lower-tier value from \{g,s,r,d,i\}. For example,
            \begin{equation*}
                \frac{\sum_{k=1}^{K} q_t^{kmg} P_k}{C_t^{mg}} \leq  n_t^{mg} ~~\forall~ m,g,t
            \end{equation*}
 
      
     
     
        
        \item Inventory balance constraints.
            \begin{equation}
              I^{kb}_{t-1} + \mathop{\sum_{a=1}^{A} q^{kab}_{t-L^{ab}} = I^{kb}_t} + \mathop{\sum_{c=1}^{C} q_t^{kbc}}~~\forall~ k,b,t 
            \end{equation}
            Here, if b $\in$ \{g,s,r,d\}, then a := next higher-tier value from \{m,g,s,r\} and c := next lower-tier value from \{s,r,d,i\}. For example,
            \begin{equation*}
              I^{kg}_{t-1} + \mathop{\sum_{m=1}^{M} q^{kmg}_{t-L_{mg}} = I^{kg}_t} + \mathop{\sum_{s=1}^{S} q_t^{kgs}}~~~\forall~ k,g,t 
            \end{equation*}
            
            
            
            
            \begin{equation*}
               I^{ki}_{t-1} + \mathop{\sum_{d=1}^{D} q^{kdi}_{t-L_{di}} = I^{ki}_t} + \mathop{\sum_{j=1}^{J} w^{ijk}_{t}}~~\forall~ k,i,t
            \end{equation*}
         
        \item Inventory Capacity Constraints.
         \begin{equation}
            \sum_{k=1}^{K} I^{ka}_t P_k \leq B^a_t ~~\forall~ a,t~~~ a \in \{g,s,r,d,i\}
        \end{equation}
        For example,
         \begin{equation*}
          \mathop{\sum_{k=1}^{K} I^{kg}_t ~P_k \leq B^g_t} ~~\forall~ g,t
        \end{equation*}
        
        \item Consumption balance constraints.
        \begin{equation}
            \mathop{\sum_{k=1}^{K} w^{ijk}_{t} + s^{ij}_{t}= \frac{D^{ij}_{t}}{w} } ~~\,\forall~ i,j,t
        \end{equation}
        
        \item Constraints on consumption incorporating lead time of administration. 
       \begin{equation}
            \mathop{\sum_{j=1}^{J} w^{ijk}_{t} \leq I^{ki}_{t-L^{ij}}} ~~\forall~ i,k,t
        \end{equation}

    \item Medical Personnel Availability Constraints.
        \begin{equation}
            \mathop{\sum_{j=1}^{J}\sum_{k=1}^{K}} T_o w^{ijk}_{t} \leq N^{i}_{t} n^{i}_{t} ~~\forall~ i,t
        \end{equation}
        
    \item Health Workers Balance Constraints.
        \begin{equation}
            n^i_t = n^i_{t-1} + h^i_t - f^i_t ~~\forall~ i,t
        \end{equation}

    \end{itemize}

%% file: Tables/Initial_parameters.tex
\begin{table}
\centering
\caption{Initial data required for estimation of capacities. \textsuperscript{1}The dimensions are estimated based on the commercially available trucks and cold boxes in India taken from \href{ https://www.indiamart.com/}{https://www.indiamart.com/}  }
\label{table:Initial_parameters1}
\footnotesize
\begin{tabular}{|l|  c| }
\hline
\multicolumn{2}{|c|}{No of CCE units in Bihar \citep{NHM}}       \\ \hline
WIF (Walk in Freezer)                 & 3             \\ \hline
WIC (Walk in Cooler)                  & 17            \\ \hline
ILR (In-line Refrigerator)            & 1039          \\ \hline
DF (Deep Freezer)                     & 887           \\ \hline
\multicolumn{2}{|c|}{Miscellaneous \citep{world2017calculate} }                   \\\hline
Utilization factor                    & 0.67          \\ \hline
\multicolumn{2}{|c|}{Standard dimensions of vehicle storage areas\textsuperscript{1} }     \\ \hline
Length (cm)             & 244           \\ \hline
Breadth (cm)           & 229           \\ \hline
Height (cm)          & 259           \\ \hline
\multicolumn{2}{|c|}{Standard dimensions of cold boxes\textsuperscript{1} } \\ \hline
Length (cm)       & 54.4          \\ \hline
Breadth (cm)           & 44.5          \\ \hline
Height (cm)             & 42            \\ \hline
\end{tabular}
\end{table}

%% file: Sections/Multi_commodity_results.tex
As done for the single commodity model, we begin our analysis for the multiple vaccine model with two cases: first with lower packed volumes per dose for each vaccine considered in the model, and a second case with higher packed volumes per dose for each vaccine. We consider two vaccines, which differ in their packed vaccine volume per dose, efficacy and cost of administration. Assuming sufficient manufacturing capacity, lead times of 1 week for delivery to the clinics from the DVS and preparation/administration at the clinics, we run the model for two sets of packed volumes:
\begin{itemize}
    \item Lower packed volume per dose. 
    \subitem Vaccine 1: a packed volume per dose of 0.211 cm\textsuperscript{3}/dose, efficacy 93.7\%, and a cost per dose of INR 780.
    \subitem Vaccine 2: a packed volume per dose of 0.086 cm\textsuperscript{3}/dose, efficacy 77.8\%, and a cost per dose of INR 1,410.
    
    \item Higher packed volume per dose. 
    \subitem Vaccine 1: a packed volume per dose of 3.3 cm\textsuperscript{3} per dose, efficacy 93.7\%, and cost per dose of INR 780. 
    \subitem Vaccine 2: a packed volume per dose of 1.719 cm\textsuperscript{3} per dose, efficacy 77.8\%, and cost per dose of INR 1410.
\end{itemize}

We observe that, as expected, with sufficient manufacturer capacity for both vaccines, the optimal solution from the multiple vaccine model reduces to that obtained from the single vaccine model. This is because all other parameter values -  such as ordering, inventory holding, and fixed and variable transportation costs - are assumed to be the same for both vaccines, and the difference in the costs per dose of the vaccines is not high enough compared to the shortage cost (as is likely to be the case) to prevent the higher efficacy vaccine to be administered. Thus we see that the higher efficacy - to be precise, the vaccine with the higher shortage cost - is selected for administration to the recipients.

In the current multiple vaccine formulation, we have used only a single efficacy parameter for the vaccines. However, vaccine efficacy may be measured with respect to multiple clinical endpoints: for example, efficacy in preventing symptomatic disease, in reducing transmission, in preventing severe disease and/or hospitalization, or in preventing death. Costs can be associated with each of these endpoints to arrive at a more comprehensive measure of vaccine efficacy. Further, some vaccines can be significantly more expensive to store or transport, and even ordering costs can be significantly higher, especially if both domestically manufactured vaccines and imported vaccines are considered for the same group of recipients. While we reserve the consideration of these complexities for future research, our work provides a proof-of-concept for how these considerations can be integrated within a decision support framework for optimizing vaccine distribution across the cold chain network.

\begin{figure}[h]
    \centering
    \includegraphics[width=5in]{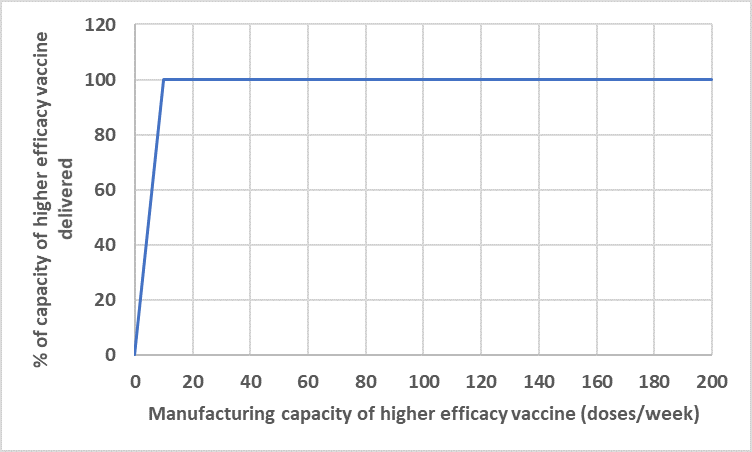}
    \caption{Proportion of the manufac higher efficacy vaccine administered among the total number of vaccine doses administered as a function of the manufacturing capacity of the higher efficacy vaccine.}
    \label{fig:Percentage of the capacity of higher efficacy vaccine  administered vs manufacturing capacity of the manufacturer of higher efficacy vaccine}
\end{figure}

We now illustrate how the multiple vaccine model can be used to determine the conditions under which the higher efficacy vaccine is not administered. We begin by analyzing the impact of manufacturer capacity. It can be seen from Figure \ref{fig:Percentage of the capacity of higher efficacy vaccine  administered vs manufacturing capacity of the manufacturer of higher efficacy vaccine} that even at low manufacturing capacities, the higher efficacy vaccine is delivered to its full capacity - that is, if the manufacturer is able to deliver 100 doses at the beginning of week $t$, all 100 doses are delivered. The remainder of the demand is fulfilled by the lower efficacy vaccine. We see that it is only at extremely low values of the manufacturer's capacity (above 10 units per week) that minimizing the costs of ordering two vaccines are prioritized lower than minimizing the shortage costs, and hence the lower efficacy vaccine units are also ordered to meet the demand requirement. It is unlikely that the situation depicted in this graph is likely to realistically occur at the figures observed; however, depending upon the differences in ordering, transportation and storage costs, such a situation may come to pass at significantly higher manufacturer capacities. This analysis provides an illustration of how our model can be used to determine the threshold capacity of the manufacturer of the higher dose vaccine below which the lower efficacy vaccines are also selected for administration.

\begin{figure}[h]
     \centering
     \begin{subfigure}[b]{0.45\textwidth}
         \centering
         \includegraphics[width=\textwidth, height=0.6\textwidth]{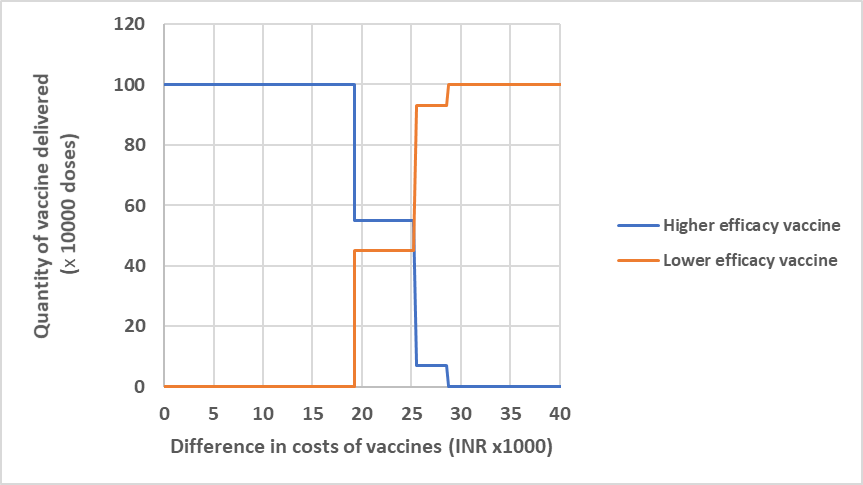}
         \caption{Impact of differences in vaccine costs per dose on quantities of each vaccine administered.}
         \label{fig:Quantity of each vaccine delivered vs Difference of cost of vaccines}
     \end{subfigure}
     \hfill
     \begin{subfigure}[b]{0.45\textwidth}
         \centering
         \includegraphics[width=\textwidth]{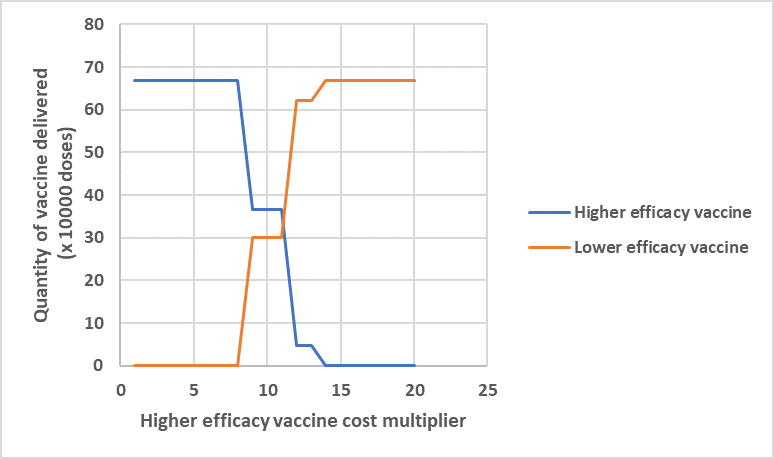}
         \caption{Impact of higher efficacy vaccine costs multiplier on quantities of each vaccine delivered}
         \label{fig:Quantity of each vaccine delivered vs multiplier of high efficacy vaccine costs}
     \end{subfigure}
        \caption{Impact of vaccine costs on the quantities of each vaccine delivered}
        \label{fig:Impact of vaccine costs on the quantities of each vaccine delivered}
\end{figure}

We now investigate the impact of costs per dose of the vaccines considered in this model. While the vaccine with higher efficacy yields a lower shortage cost, its cost per dose could be greater than that of the vaccine with lower efficacy. The multiple vaccine model can be used to determine the difference in the costs of the two vaccines at which the vaccine with the lower efficacy administered, given the sufficient manufacturer capacity for both the vaccines. We illustrate this now. In this analysis, we keep all the costs for the two vaccines same except for their shortage costs and costs per dose. We see from Figure \ref{fig:Quantity of each vaccine delivered vs Difference of cost of vaccines} that when the difference between the cost of vaccine is less than INR 19,300, it is always the higher efficacy vaccine that is administered but beyond this, some doses of the lower efficacy vaccine are also administered.

We see that once this threshold difference in costs per dose between the vaccines is breached, the cost of the higher efficacy vaccine dominates the shortage cost for the subgroup comprising children, but does not do so for the adult and elderly subgroups due to their higher shortage costs. Hence, the demand for the subgroup comprising children is met by the lower efficacy vaccine while the other subgroups receive the higher efficacy vaccine. When the cost per dose difference exceeds INR 25,200, adults also receive the lower efficacy vaccine, and when it exceeds INR 28,600, all recipients receive the lower efficacy vaccine. 

Once again, we note that while it is very unlikely that vaccines will differ in their costs per dose by as much as INR 19,300, the lower efficacy vaccine may be preferred if differences in other associated costs are also significantly higher for the higher efficacy vaccine, and the vaccine efficacies are not substantially different. This is illustrated in Figure~\ref{fig:Impact of vaccine costs on the quantities of each vaccine delivered}. In this analysis, we increased all fixed and variable costs that could conceivably be greater for the higher efficacy vaccine (e.g., the cost per dose, the fixed and variable transportation costs, inventory holding, and fixed ordering costs), and we find that when these costs are 9 times that of the lower efficacy vaccine, the lower efficacy vaccine starts receiving orders. As the vaccines become closer in efficacy, the value of this multiplier will also accordingly decrease. 

We now consider the computational expense of the single vaccine model within our framework, and describe methods to reduce runtimes for various types of analyses.